\documentclass[aps,prd,showpacs,twocolumn,superscriptaddress,nofootinbib,preprintnumbers]{revtex4-2} 

\makeatletter
\def\p@subsection{}
\makeatother

\usepackage{graphicx,amsmath,amssymb,lipsum,bm}

\usepackage[dvipsnames]{xcolor}
\definecolor{xlinkcolor}{rgb}{0.7752941176470588, 0.22078431372549023, 0.2262745098039215}
\definecolor{BrickRed}{rgb}{0.7752941176470588, 0.22078431372549023, 0.2262745098039215}
\definecolor{xlinkcolor}{HTML}{1c1e94}
\usepackage{booktabs} 
\usepackage[normalem]{ulem}
\usepackage[utf8]{inputenc} 
\usepackage{mathtools}
\usepackage{aas_macro}
\usepackage{xspace}

\usepackage{ulem}
\usepackage{diagbox}

\usepackage[dvipsnames]{xcolor}
\usepackage{mathrsfs}

\usepackage[colorlinks=true,citecolor=xlinkcolor,linkcolor=xlinkcolor,urlcolor=xlinkcolor, backref=false,pdfborder={0 0 0}]{hyperref}

\newcommand{\beqa}{\begin{eqnarray}}
\newcommand{\eeqa}{\end{eqnarray}}

\newcommand{\be}{\begin{equation}}
\newcommand{\ee}{\end{equation}}
\newcommand{\beq}{\begin{equation}}
\newcommand{\eeq}{\end{equation}}

\newcommand\p{{\bm p}}
\renewcommand\k{{\bm k}}
\newcommand\q{\bm{q}}

\newcommand{\bseq}{\begin{subequations}}
\newcommand{\eseq}{\end{subequations}}

\def\ltsima{$\; \buildrel < \over \sim \;$\xspace}
\def\gtsima{$\; \buildrel > \over \sim \;$\xspace}
\def\simlt{\lower.5ex\hbox{\ltsima}}
\def\simgt{\lower.5ex\hbox{\gtsima}}

\newcommand{\hinvMpc}{\,h^{-1}\, {\rm Mpc}\xspace}

\newcommand{\hMpcinv}{\,h\, {\rm Mpc}^{-1}\xspace}

\newcommand{\hMpc}{\, h\mathrm{Mpc}^{-1}\, }

\newcommand{\Lya}{Ly-$\alpha$\xspace}
\newcommand{\td}{\delta}
\newcommand{\dd}{\partial}

\newcommand{\vk}{\mathbf{k}}

\def\gsim{\raise0.3ex\hbox{$\;>$\kern-0.75em\raise-1.1ex\hbox{$\sim\;$}}}
\def\lsim{\raise0.3ex\hbox{$\;<$\kern-0.75em\raise-1.1ex\hbox{$\sim\;$}}}

\def\beqn#1{\begin{equation}\label{#1}}
\def\eeqn{\end{equation}}

\def\beqa#1{\begin{eqnarray}\label{#1}}
\def\eeqa{\end{eqnarray}}

\newcommand{\abacus}{\textsc{AbacusSummit}\xspace}

\def\hMpc{h{\text{Mpc}}^{-1}}

\def\Z2{$\mathcal{Z_2}$}

\def\vpsi{{\boldsymbol{\psi}}}

\newcommand {\ignore}[1]{}

\DeclareRobustCommand{\ion}[2]{%
\relax\ifmmode
\ifx\testbx\f@series
{\mathbf{#1\,\mathsc{#2}}}\else
{\mathrm{#1\,\mathsc{#2}}}\fi
\else\textup{#1\,{\mdseries\textsc{#2}}}%
\fi}

\newcommand{\BCCP}{Berkeley Center for Cosmological Physics, Department of Physics, University of California, Berkeley, CA 94720, USA}
\newcommand{\LBL}{Lawrence Berkeley National Laboratory, One Cyclotron Road, Berkeley CA 94720, USA}
\newcommand{\MIT}{Center for Theoretical Physics -- a Leinweber Institute, Massachusetts Institute of Technology, Cambridge, MA 02139, USA}
\newcommand{\IAIFI}{The NSF AI Institute for Artificial Intelligence and Fundamental Interactions, Cambridge, MA 02139, USA}

\begin{document}

\preprint{MIT-CTP/5882}

\title{Modeling the Cosmological Lyman--$\alpha$ Forest at the Field Level}

\author{Roger de Belsunce}
\email{rbelsunce@lbl.gov}
\affiliation{\LBL}
\affiliation{\BCCP}
\author{Mikhail M. Ivanov}
\email{ivanov99@mit.edu}
\affiliation{\MIT}
\affiliation{\IAIFI}
\author{James M.~Sullivan}
\email{jms3@mit.edu}\thanks{Brinson Prize Fellow}
\affiliation{\MIT}
\author{Kazuyuki Akitsu}
\affiliation{Theory Center, Institute of Particle and Nuclear Studies, High Energy Accelerator Research Organization (KEK), Tsukuba, Ibaraki 305-0801, Japan}
\author{Shi-Fan Chen}
\affiliation{Institute for Advanced Study, 1 Einstein Drive, Princeton, NJ 08540, USA}

\begin{abstract} 
The distribution of 
absorption lines 
in the spectra of distant quasars, called the Lyman-$\alpha$ (\Lya) 
forest, is
a unique probe of cosmology and the
intergalactic medium at high 
redshifts and small scales. 
The statistical power
of ongoing 
redshift surveys demands
precise theoretical tools 
to model the \Lya forest.
We address this challenge by
developing an analytic, perturbative 
forward model to predict the  
\Lya
forest at the field level
for a given set of cosmological initial conditions. 
Our model shows a remarkable 
performance when compared with 
the Sherwood hydrodynamic simulations: it reproduces the flux distribution, the
\Lya--dark matter halo cross-correlations,
and the count-in-cell statistics 
at the percent level
down to scales of a few Mpc.
Our work provides crucial 
tools 
that bridge analytic modeling
on large scales
with simulations on small-scales,
enabling field-level inference 
from \Lya\ forest data and  
simulation-based priors
for cosmological analyses.    
This is especially timely for realizing the full scientific potential 
of the \Lya forest measurements 
by the Dark Energy Spectroscopic Instrument.
\end{abstract}

\maketitle

\textit{Introduction.}---Fluctuations in the cosmological density field encode information about the composition and evolution of the Universe. While galaxies, dark matter halos, and quasars trace the highly overdense regions, the low-density regions are filled with diffuse gas. In particular, neutral hydrogen in the low-density, highly ionized intergalactic medium (IGM) absorbs light, producing a distribution of absorption lines in observed quasar spectra, known as the Lyman-$\alpha$ (\Lya) forest. Measurements of these fluctuations in the neutral hydrogen density provide a powerful probe of the large-scale structure  of the Universe in the high-redshift regime ($2\simlt z \simlt 4$) as well as the thermal and ionization state of the IGM (see, e.g.,~\cite{McQuinn:2016} for a review).

The large cross-section of the \Lya absorption of neutral hydrogen makes this observable sensitive down to Mpc distances and below,
making it a unique 
source of cosmological
information on small scales. 
In particular, the measurements of the line-of-sight power spectrum
of the 
\Lya forest (see, e.g.,~\cite{Seljak:2005,Viel:2005,McDonald06,PYB13,Chabanier:2019,Pedersen:2020,2023MNRAS.526.5118R,2024MNRAS.tmp..176K}) constrain neutrino physics \cite{Seljak:2005,Viel:2010,PYB13,Palanque2020, Ivanov:2024jtl, He:2023oke, He:2025jwp}, primordial black holes \cite{Afshordi:2003,Murgia:2019}, dark matter \cite{Viel:2013,Baur:2016,Irsic17,Kobayashi:2017,Armengaud:2017,Murgia:2018,Garzilli:2019,Irsic:2020,Rogers:2022,Villasenor:2023,Irsic:2023}, the thermal state of the ionized IGM \cite{Zaldarriaga:2002,Meiksin:2009,McQuinn:2016,Viel:2006,Walther:2019,Bolton:2008,Garzilli:2012,Gaikwad:2019,Boera:2019,Gaikwad:2021,Wilson:2022,Villasenor:2022}, non-minimal cosmological models \cite{Goldstein:2023gnw,
Garny:2018byk,
Fuss:2022zyt}, and the running of the spectral index \cite{Seljak:2006bg,Ivanov:2024jtl}.
Combining multiple lines of sight allows
one to reconstruct three-dimensional 
(3D) 
cosmological 
fluctuations on large scales.
These 3D
auto correlations of the 
forest and cross-correlations
between the forest and quasars 
constrain the expansion history of our Universe through measurements of the baryon acoustic oscillations (BAO; \cite{McDonald:2007,Slosar2013,Busca:2013,dMdB:2020}), 
and the broadband shape of the 3D correlation function \cite{Slosar2013,Cuceu:2021,Cuceu:2023,Gordon:2023}. 
The \Lya forest at high redshifts plays a crucial
role in strengthening recent evidence for dynamical dark energy from the Dark Energy Spectroscopic Instrument (DESI)
experiment~\cite{DESI:2025zgx}. 

DESI~\cite{DESI:2016,DESI:2022, DESI_BAO_2024,DESI_lya_2024} will observe approximately one million 
quasar spectra 
with \Lya forest
over its five-year observation period, producing the 
most precise map of our Universe at high redshift.
Next-generation surveys such as DESI-II and Spec-S5 \cite{Schlegel:22_spec_roadmap,Besuner25:spec_s5} 
will extract the 
\Lya forest from spectra
of high-redshift 
galaxies, which will 
increase the 
total cosmological signal 
even further. 
These impressive 
experimental achievements
call for robust theoretical 
modeling tools whose precision has to match 
statistical errors 
in order to convert 
the ongoing and upcoming 
\Lya measurements 
into precision measurements 
of cosmological parameters. 

Current \Lya forest modeling capabilities do not rise to the challenge posed by the incoming quantity and quality of this data.
To date, all analyses of the three-dimensional
\Lya forest data have been done at the level
of the two-point function \cite{DESI_BAO_2024} (or its Fourier counterpart, the power spectrum \cite{Font-Ribera:2018, deBelsunce:2024knf}).
While the two-point
function analysis
is computationally 
manageable and 
easy to interpret,
the compression
of a three-dimensional field
into the two-point function 
is inevitably
lossy~\cite{Nguyen:2024yth,Akitsu_FLI}. 
A more optimal analysis 
may be a
field level inference
or the inclusion of
three- and four-point
function correlations.
All these methods
require consistent modeling of the \Lya forest beyond the two-point function. 
While such modeling is, in theory, 
possible by means of numerical hydrodynamic simulations, this method
cannot be easily scaled to reproduce large 
cosmological volumes relevant to 
cosmic surveys.\footnote{For the analysis of large scales, different 
approximate
prescriptions have been used to create large-volume \Lya forest mocks: \abacus $N$-body simulations paint the forest on top of the dark matter field \cite{Hadzhiyska:2023, Hadzhiyska:2025cvk} using a fluctuating Gunn-Peterson approximation (FGPA) calibrated on hydrodynamic simulations \cite{Croft98,2022ApJ...930..109Q}. More sophisticated techniques include the \Lya Mass Association Scheme (LyMAS; \cite{Peirani:2014,Peirani:2022}), the Iteratively Matched Statistics method (IMS; \cite{Sorini:2016}), Hydro-BAM \citep{2022ApJ...927..230S}, cosmic-web-dependent FGPA \citep{2024A&A...682A..21S}, and deep-learning reconstruction \cite{Jacobus:2024yev} 
to connect the observed flux to the underlying matter field.
} 
In this \textit{Letter},
we provide the theoretical tools
for field-level modeling 
of the \Lya forest applicable to arbitrarily large volumes, 
enabling 
analysis modes and methodologies
that have not been possible before.

We develop a forward model
for the \Lya forest flux fluctuations at the field level based on 
the framework of effective field theory (EFT;~\cite{McDonald:2009dh,Baumann:2010tm,Carrasco:2013mua,Ivanov:2022mrd}).
The use of EFT is crucial 
for our purpose because 
it provides a systematic
first-principles theoretical
description of the \Lya field
on large scales based
on symmetries and dimensional
analysis only.
Conceptually, our 
model is similar to the 
field-level forward model for 
galaxies~\cite{Schmittfull:2020trd, Obuljen:2022cjo},
but it includes the line-of-sight
dependent operators dictated by symmetries of the \Lya forest~\cite{McDonald:1999dt,Givans:2020sez,Desjacques:2018pfv,Chen:2021rnb,Ivanov:2023yla,Ivanov:2024jtl, Belsunce_Sullivan_skewspectrum}. 
We test our model against
the high-fidelity hydrodynamic Sherwood simulation \cite{Bolton:2016bfs}, 
and find that it can accurately 
reproduce the simulated
\Lya forest distribution on scales
greater than a few Mpc.
Crucially, in order to match the two
realizations, the amplitudes and phases of all Fourier modes
have to be correct.
We show that our forward model can match \emph{all} the amplitudes and phases of \emph{all} Fourier modes, thus passing a more stringent accuracy test than comparing summary statistics which average over modes.

\textit{Forward model.}---The quantity of interest is
the fluctuation field of the \Lya forest transmitted flux $ \td_F = F/\overline{F}(z)-1$ around the mean value of transmission $\overline{F}(z)$,
which depends on redshift $z$.
In what follows we aim to model 
the \Lya field from a hydrodynamical 
simulation, which we will refer to as a ``true''
field $\delta^{\rm truth}_F$.
On large scales, 
the equivalence principle demands that
the flux fluctuations 
trace the underlying dark matter
overderdensity $\delta$ 
and its line-of-sight velocity gradient~$\eta = \dd_{\parallel}v_\parallel/(aH)$ ~\cite{McDonald:1999dt,McDonald:2001} 
\beq 
\label{eq:lint}
\td^{\rm truth}_F \approx \td^{\rm model}_F=b_1\td - b_{\eta}\eta \,, 
\eeq 
where $a$ and $H=\dot a/a$ are the metric scale factor and the Hubble parameter, respectively, 
while $b_1$ and $b_{\eta}$
are the linear bias parameters
of the \Lya forest. 
The above formula 
produces the 
well-known linear-level flux power spectrum (equivalent to the Kaiser formula for galaxies \cite{Kaiser:1987qv}):
\beq 
\label{eq:kais}
P^{\rm lin}_F(k,\mu) = (b_1 - b_{\eta}f\mu^2)^{2} P_{\rm lin}(k)\,,
\eeq 
where $k$ is the Fourier wavenumber, $\mu= k_\parallel / k$ the cosine of the angle to the line-of-sight and the expression is evaluated at the
effective redshift of the forest $z$ (here: redshift of the simulation snapshot) and in what follows
we suppress the explicit
time-dependence. 
In linear theory, 
the model in Eq.~\eqref{eq:lint}
is equivalent to
\beq 
\label{eq:lint_2}
\td^{\rm model}_F(\k) = \beta_1(k,\mu)\td_1(\k) \,, 
\eeq 
where $\delta_1$ is the linear matter field and $\beta_1(k,\mu)$
is the momentum-dependent transfer function \cite{Schmittfull:2018yuk,Schmittfull:2020trd}. On large scales it should approach 
the perturbative value $\beta_1^{\rm linear}\equiv b_1-b_\eta f\mu^2$ consistent with Eq.~\eqref{eq:kais}.

Eq.~\eqref{eq:lint}, however, 
is not expected to be perfectly 
accurate even on large scales
as the distribution 
of the \Lya flux is 
not completely deterministic. 
The error of the linear model
is captured by the stochastic field $\epsilon$:
\beq 
\label{eq:lint_eps} 
\td^{\rm truth}_F(\k) = \beta_1(k,\mu)\td_1(\k)+\epsilon(\k), 
\eeq 
Given
$\td^{\rm truth}_F$, the 
power
spectrum of $\epsilon$ can be computed by 
minimizing the mean-square 
difference $\langle |\delta_F-\beta\td_1|^2 \rangle$ in each wavenumber bin producing the best possible perturbative model provided that 
\be 
\label{eq:tr1}
\beta_1(k,\mu)=\frac{\langle 
\delta_F^{\rm truth}(\k)
\delta^*_1(\k)\rangle}{\langle 
|\delta_1(\k)|^2\rangle}\,.
\ee 
A significant scale-dependent
departure from $\beta_1^{\rm linear}$ implies that higher order corrections must be included. 

The power spectrum of $\epsilon$,
refereed to as 
the error (or noise)
power spectrum 
\be 
P_{\rm err} (k,\mu) \equiv \langle |\delta^{\mathrm{truth}}_F(\k) - \delta^{\mathrm{model}}_F(\k)|^2 \rangle\,,
\ee 
reflects the agreement at the level of the phases.
As such, it 
is a measure
of the success of the model:
a large, scale-dependent error spectrum implies that the model performs poorly. 
The noise power spectrum is related to the cross-correlation coefficient between the true 
field and the model, $\delta_F^{\rm model}$,
via $P_{\rm err}=P_{\rm truth}(1-r^2_{cc})$,
where 
\be 
r_{cc}(\delta_F^{\rm truth},\delta^{\rm model}_F)=\frac{\langle 
\delta^{\rm model}_F(\k)
[\delta_F^{\rm truth}(\k)]^*
\rangle}{
\langle 
(|\delta_F^{\rm truth}(\k)|^2\rangle
\langle
|\delta_F^{\rm model}(\k)|^2\rangle)^{1/2}
}\,.
\ee 
EFT predicts that once all deterministic 
terms are included in the model~\cite{Ivanov:2023yla}, 
\be \label{eq:perr_ns}
P_{\rm err}=n_0(1+\alpha_1 k^2+
\alpha_2 
\mu^2 k^2)\quad \text{as}\quad k\to 0\,,
\ee 
where
$n_0,\alpha_1,\alpha_2$ are dimensional constants.
A successful forward model should re-produce
the above behavior with 
small $n_0$, $\alpha_{1,2}$, and 
a weak scale dependence. 
For galaxies, $n_0$ is set by the number density. 
In analogy with galaxies, in the \Lya context $n_0$
is believed to be small
due to the large number density
of absorption lines
per spectrum. 
Given that this noise
is a crucial limiting 
factor for
cosmological inference,
it is important to quantify
it accurately, 
which we do 
for the first time in 
this \textit{Letter}.

Beyond linear theory,
the correlation between the \Lya forest and the underlying dark matter requires introducing  line-of-sight dependent operators \cite{McDonald:1999dt,McDonald:2001,Chen:2021,Ivanov:2023yla}, whose most general form is 
\be
\label{eq:general_expansion}
\delta_F(\vk, z) = \sum_{\mathcal{O}}b_{\mathcal{O}}(z)\mathcal{O}(\vk,z) + \epsilon(\vk, z)\,,
\ee
which consists of a deterministic contribution stemming from the bias expansion for each bias operator, $\mathcal{O}$, with bias parameters $b_{\mathcal{O}}$.

Replacing the bias operators with $k-$dependent transfer functions leads to a flexible model for the \Lya forest field that reduces to the perturbative bias expansion on large scales. To reproduce
EFT to cubic order
it is sufficient
to use 
operators through the second order supplemented with 
appropriate transfer functions~\cite{Schmittfull:2018yuk,Schmittfull:2020trd}.
To eliminate redundancy among the transfer functions, we orthogonalize the set of operators in Eq.~\eqref{eq:general_expansion} using the Gram-Schmidt algorithm~\cite{Schmittfull:2018yuk}, and obtain
\begin{align}
\label{eqn:lya_model}
   & \td^{\rm model}_F(\vk) = 
   \beta^F_1(k,\mu)\tilde \delta_1(\vk)\\
   &
   +\beta^F_{\eta}(k,\mu)\left(
   \delta_Z(\vk)-\frac{3}{7} f \mu^2\tilde{\mathcal{G}_2} \right)^\perp\nonumber \\
&+\beta^F_2(k,\mu)
\tilde{(\delta_1^2)}^\perp (\vk) +\beta^F_{\mathcal{G}_2}(k,\mu)
\tilde{\mathcal{G}_2}^{\perp}(\vk) \nonumber\\
&
+\beta^F_{\delta \eta}(k,\mu)
\tilde{[\delta \eta]}^\perp (\vk)\nonumber\\ &+\beta^F_{\eta^2}(k,\mu)
\tilde{\eta}^{2,\, \perp} (\vk)  
+\beta^F_{KK_\parallel}(k,\mu)
\tilde{(KK)}_\parallel^\perp (\vk)\,, \nonumber 
\end{align}
where the operators are given in the supplemental material.\footnote{Note that we do not include the transfer function associated with $\Pi_\parallel^{[2]}$ as it is fully degenerate with the other operators, see Supplemental Material.} 
To account for the non-perturbative
dependence on the 
linear displacements relevant 
for the BAO (i.e. IR resummation~\cite{Senatore:2014via,Baldauf:2015xfa,Vlah:2015zda,Blas:2015qsi,Blas:2016sfa,Ivanov:2018gjr,Vasudevan:2019ewf,Hadzhiyska:2025cvk}), 
we shift the operators above
by the Zel'dovich displacements.\footnote{While the simulation we use
does not resolve the BAO because of the small volume, 
including IR resummation
is important as it produces
unique operators in the \Lya that
which are 
protected by the equivalence 
principle. } 
The resulting EFT cubic bias model 
is formally equivalent to the one-loop
EFT 
power spectrum \cite{Ivanov:2023yla}
and tree-level bispectrum~\cite{Belsunce_Sullivan_skewspectrum} with the transfer functions absorbing  further
higher-order deterministic 
contributions. 

In what follows we will 
compare the \Lya field
produced with the model
in Eq.~\eqref{eqn:lya_model}
evaluated for the 
true initial dark matter
density field realization $\delta_1$,
against the simulated
field, $\delta_F^{\rm truth}$.
In that case the optimal
transfer functions 
that minimize $P_{\rm err}$
are given by the 
cross-spectra similar to 
Eq.~\eqref{eq:tr1}~\cite{Schmittfull:2018yuk,Schmittfull:2020trd}.
These transfer functions can be either computed
in EFT~\cite{Ivanov:2024xgb}
or fitted 
from data
with polynomials
in $k$, $\mu$~\cite{Obuljen:2022cjo}.
Once the transfer functions 
are calibrated, 
they can be applied to 
other realizations $\delta_1$.
The universality of EFT 
guarantees that the large-scale
distribution of the resulting \Lya field
will be indistinguishable 
from a full hydrodynamical
simulation run with the 
same initial field realization.

\textit{Data.}---We validate our perturbative forward model on the hydrodynamic simulations of the intergalactic medium (IGM) from the Sherwood suite \cite{Bolton:2016bfs,Givans2022}.
We use the largest available box with a comoving volume $V=160^3(\hinvMpc)^3$ and $2048^3$ particles. Both, the perturbative model and the simulations use the same initial conditions which are generated using the \texttt{N-GenIC} code \cite{Springel05} based on the best-fit \emph{Planck} cosmology with cosmological parameters set to $\Omega_m=0.308,\, \Omega_b=0.0482,\, h=0.678,\, \sigma_8=0.829,\, n_s=0.961$ \cite{Planck_first}.\footnote{For more information we refer the reader to \cite{Givans2022}.} The simulation snapshot is centered at redshift $z=2.8$ with a grid resolution of 1024 cells in the $x,y$ directions and of 2048 along the line of sight, chosen to be the $z$-axis throughout this work. The redshift space distortions have been applied along the same line of sight.

The Sherwood simulation additionally provides a halo catalog which we use to demonstrate the ability of our forward model to consistently 
describe the cross-correlation of the \Lya forest and halos.
Massive and light halos in this
context play the role of 
proxies for quasars and high-redshift galaxies, 
respectively~\cite{Chudaykin:2025gsh}. 

\begin{figure*}
    \centering
    \includegraphics[width=0.49\linewidth]{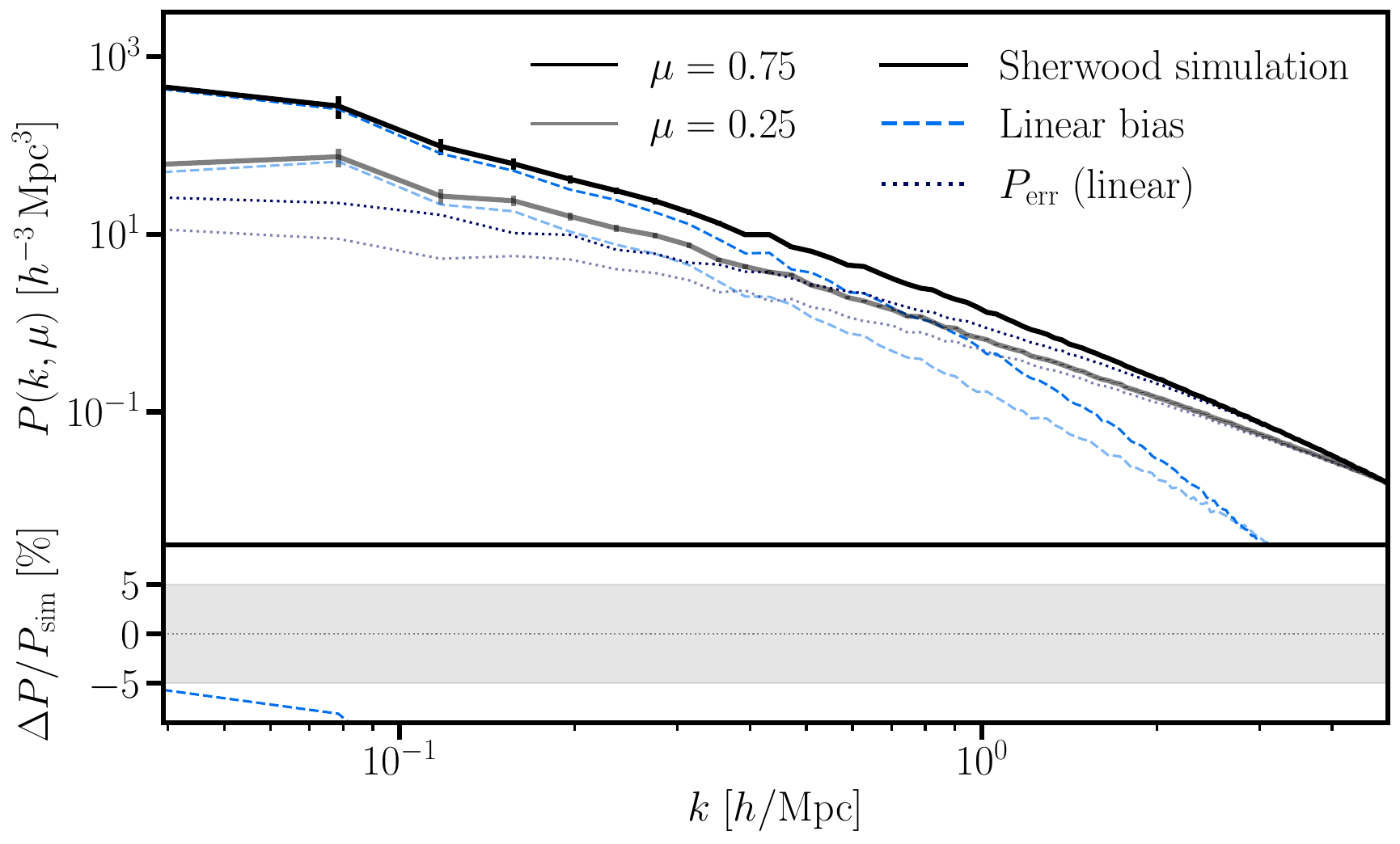}\hfill
    \includegraphics[width=0.49\linewidth]{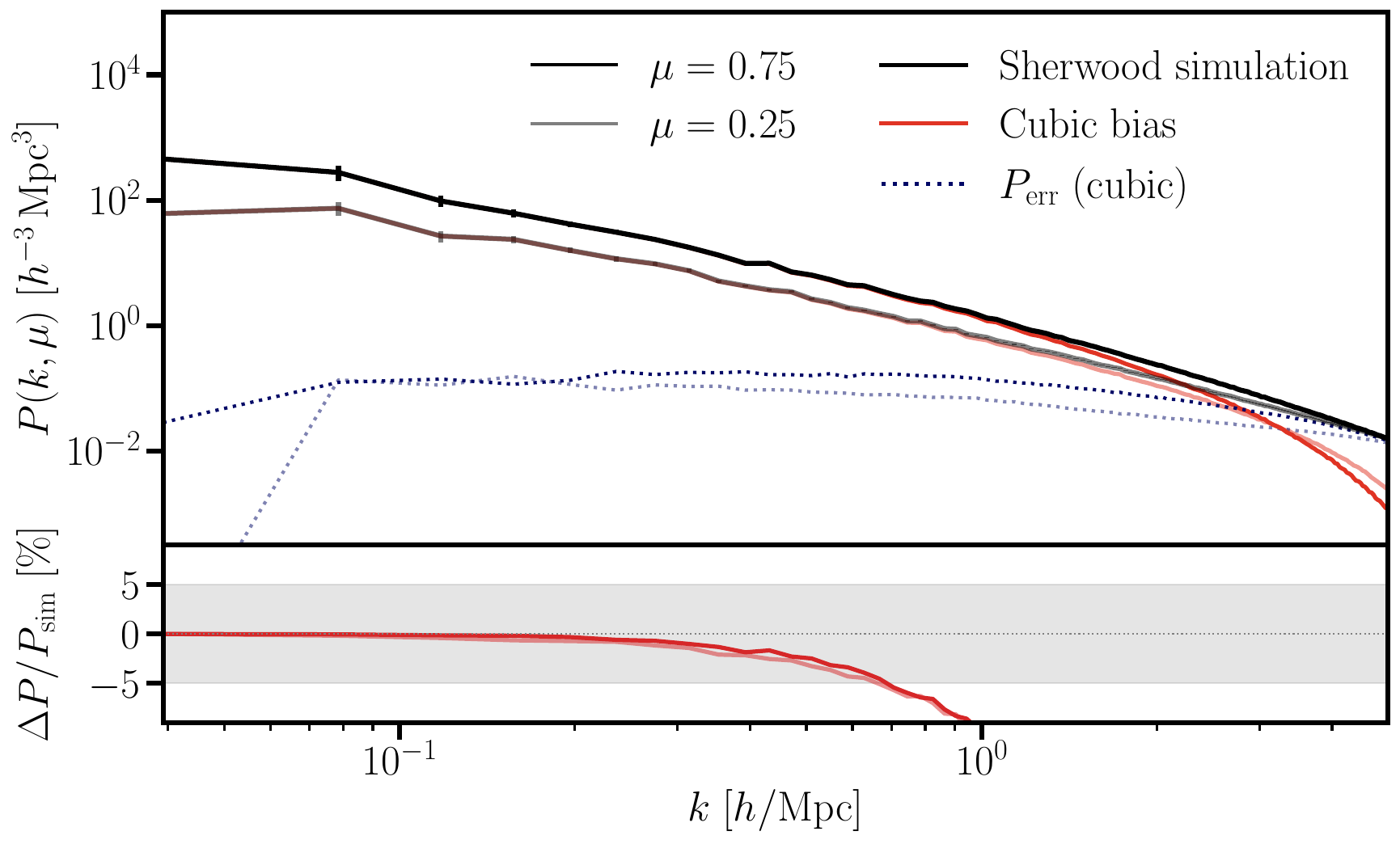}
    
    \vspace{-0.1in}
    \caption{
    Comparison between the measured two-dimensional power spectrum from the Sherwood simulation (black) and the best-fit forward model obtained from linear and full EFT cubic bias models. \textit{Left}: Linear theory prediction (light blue dashed line), which shows significant deviations from the simulation across all scales and a large error power spectrum $P_{\mathrm{err}}(k,\mu) \equiv \langle |\delta^{\mathrm{truth}}_F - \delta^{\mathrm{model}}_F|^2 \rangle$ (blue dotted lines), indicating poor model performance. \textit{Right}: EFT model prediction (red solid line), which matches the simulation much more closely, with substantially reduced residuals and a suppressed $P_{\mathrm{err}}$. In each panel, the power spectrum is shown in bins of Fourier wavenumber $k$ and angle to the line of sight, parametrized by $\mu = k_\parallel / k$, with darker (lighter) lines for $\mu = 0.75$ ($0.25$). The bottom panel displays the percent difference between the simulation and model power spectra.  
    A gray band highlights the $\pm5\%$ region in the bottom panel.
    }
    \label{fig:lya_pk}
\end{figure*}
\begin{figure*}
    \centering
    \includegraphics[width=1\linewidth]{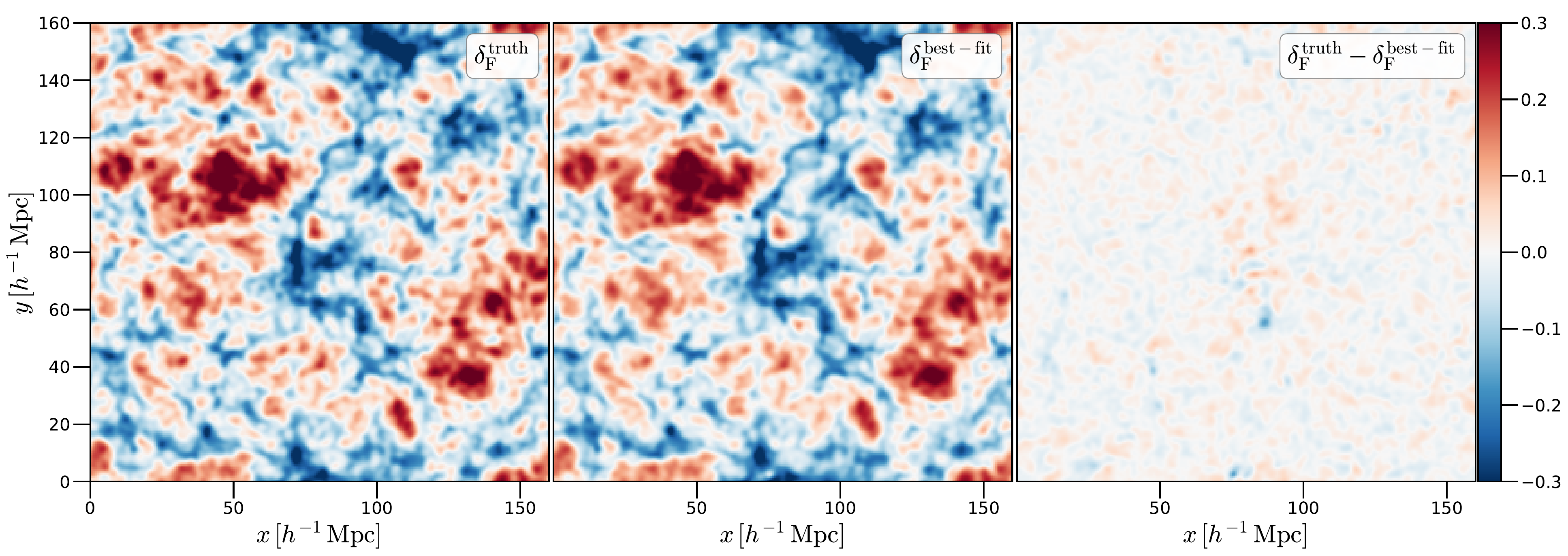}
    \includegraphics[width=1\linewidth]{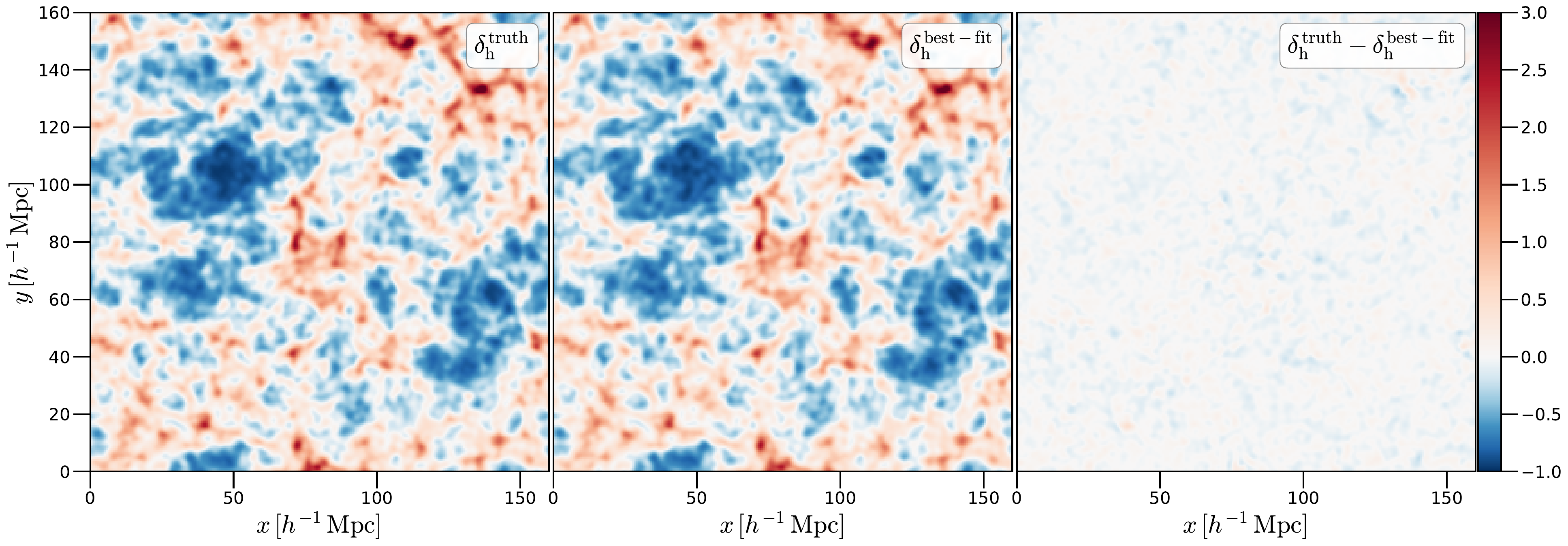}
    \caption{
    \textit{Upper panel:} 
    Comparison of slices of the flux decrement from the Sherwood hydrodynamic simulations at redshift $z=2.8$ (\textit{left panel}) to the best-fit EFT forward model (\textit{center}) and the residuals (\textit{right}). All density fields are smoothed using a three-dimensional Gaussian isotropic kernel with $R=1 \hinvMpc$. The line-of-sight is chosen to be the $z$-axis and each slice through the field is of depth $25\hinvMpc$. 
    \textit{Lower panel:} Same as above, but for the halo density field using all available halo masses.
    }
    \label{fig:lya_rsd_field}
\end{figure*}

\begin{figure*}
    \centering
\includegraphics[width=0.49\linewidth]{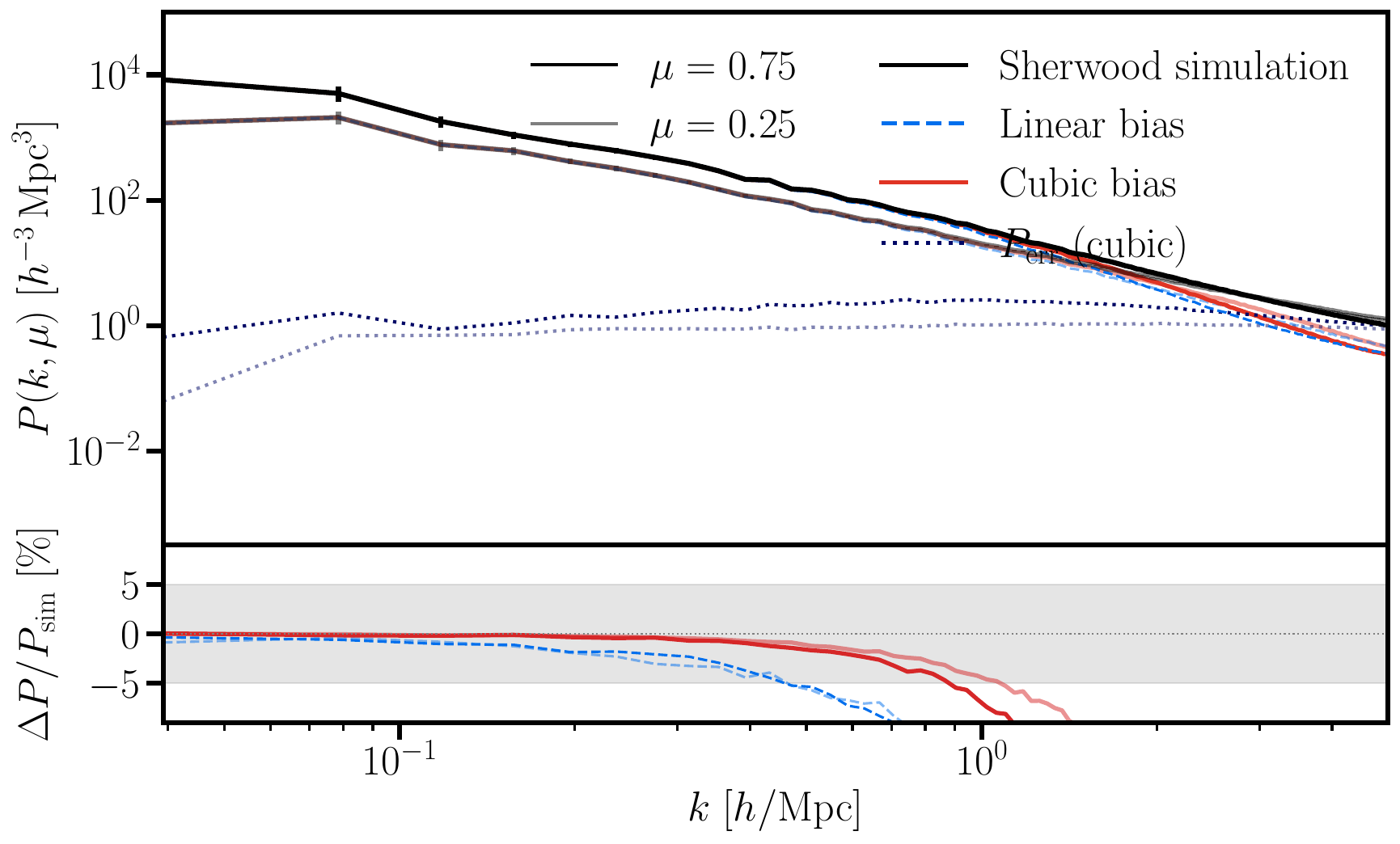}\includegraphics[width=0.49\linewidth]{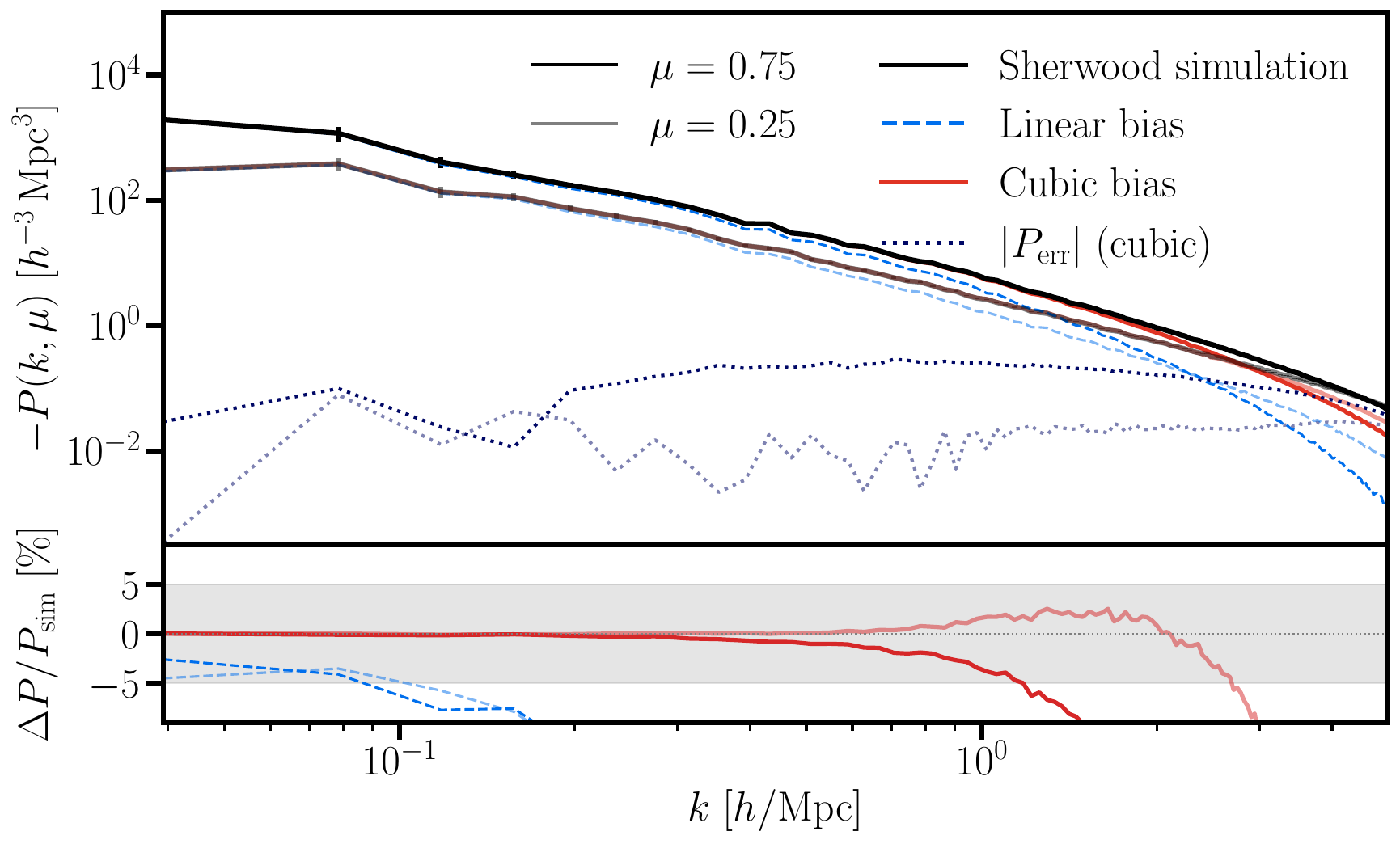}
    \caption{Comparison of the halo auto-power spectrum (\textit{left panel}) with the \Lya--halo cross-power spectrum (\textit{right panel}), constructed by combining the forward models used in Figs.~\ref{fig:lya_pk} and \ref{fig:lya_rsd_field}. 
    }
    \label{fig:lyaxqso_pk}
\end{figure*}

\textit{Results.}--
In Fig.~\ref{fig:lya_pk} we quantify the agreement between the simulations and the forward model at the power spectrum level. We use two angular bins centered at $\mu=0.25$ and 0.75. We compare the measured power spectra from the Sherwood snapshot (black lines) to the ones computed from the forward modeled fields generated using the linear model from Eq.~\eqref{eq:lint_2} (left plot; blue dashed) and the full EFT cubic bias model from Eq.~\eqref{eqn:lya_model} (right plot; red solid).
A comparison at the field-level using the one-point probability density function is presented
in Supplemental Material.
Examining the 
performance of the linear model
from Eq.~\eqref{eq:lint_eps} (left panel of Fig.~\ref{fig:lya_pk}),
we observe that linear theory 
supplemented with a transfer 
function
fails even on very large scales, 
producing a large  
scale- and orientation-dependent 
error contributing to more
than $5\%$ of the total signal
on all scales.\footnote{This result is consistent 
with earlier studies
of the \Lya at the field level~\cite{Cieplak:2015kra}.}

However, 
including
the EFT corrections
changes the reach 
of the theory dramatically,
see the right panel of Fig.~\ref{fig:lya_pk}.
 The EFT forward model $P_{\mathrm{err}}$ is orders of magnitude smaller than the deterministic part of the power spectrum
 all the way up to $k\approx 0.6 \hMpcinv$,
 where it crosses the 
 5\%-threshold.
 The noise is white on large scales, and the onset of its scale
 and orientation dependence
 is well described by the EFT predictions~\eqref{eq:perr_ns}.
 These results underscore the limitations of linear theory and the necessity of a higher-order bias expansion for an accurate forward model. This also illustrates 
 that a field-level test
 of the forward model is a much more
 stringent test than a power spectrum-based
 comparison. For instance, 
 the forward model
 $\delta_F^{\rm model}=(\frac{P_{\rm F}(k,\mu)}{P_{\rm lin}(k)})^{1/2}\delta_1(\k)$ 
 where $P_{\rm F}(k,\mu)$
 is the non-linear \Lya forest power spectrum by construction
provides a perfect match 
to the power spectrum on all scales, but fails 
at the field level beyond $0.04~\hMpcinv$
because it does not 
match the phases.\footnote{At the field level this model is equivalent to our linear model, which fails at $k\gtrsim 0.04~\hMpcinv$, see the left panel of Fig.~\ref{fig:lya_pk}.} In Supplemental Material we further quantify
the performance of the model at the level of the one-point probability density functions.

In the top row of Fig.~\ref{fig:lya_rsd_field}, we compare the Sherwood snapshot of the \Lya\ forest flux decrement field, $\td_F^{\rm truth}$ (left panel), with the corresponding field from our perturbative forward model, $\td_F^{\rm best-fit}$ (center panel), and show the residuals between the two (right panel). 
Qualitatively, the agreement between the hydrodynamic simulation and the perturbative model is excellent. Notably, the largest residuals appear in underdense regions near the center of the figure which occur in the vicinity of halos, shown in the bottom row of Fig.~\ref{fig:lya_rsd_field} in red. This observation 
provides an important 
insight into 
the \Lya 
forest stochasticity.

We also model the cross-correlation
between the \Lya forest and halos by combining our \Lya
field model 
and the halo forward
model in redshift space from~\cite{Schmittfull:2020trd}. 
The resulting 
auto-spectrum 
of halos
is shown in the left panel and the \Lya forest -- halo cross-spectrum in the right panel of Fig.~\ref{fig:lyaxqso_pk}. 
Combining both models captures the cross-spectrum of both fields at very high accuracy down to $k \simlt 1 \hMpcinv$ (c.f.~\cite{Givans:2022qgb,Chudaykin:2025gsh}). 
The excellent agreement at the field level from the bottom row of Fig.~\ref{fig:lya_rsd_field} is further discussed in Supplemental Material. 

\textit{Model error power spectrum.}---Our work provides the first estimates of the 
stochasticity of the 
3D \Lya fluctuations. 
Stochasticity is of 
great importance to 
cosmological constraints
because it determines 
the irreducible error floor
or the structure formation 
``background,'' 
which is uncorrelated with
the cosmological initial conditions. 
EFT is aimed at modeling 
the deterministic part, while for the stochastic
part only the simple 
power law momentum expansion 
is available. 
For this expansion
applied to the noise power
spectrum of the \Lya field in \eqref{eq:perr_ns} we 
find $n_0\approx 0.18~$ $[h^{-1}\text{Mpc}]^3$, $\alpha_1\approx -0.25~[h^{-1}\text{Mpc}]^{2}$,
$\alpha_2\approx 0.51~[h^{-1}\text{Mpc}]^{2}$. The  
constant \Lya ``shot noise''
$n_0$
does not follow
from any natural scale
of the \Lya physics. 
The appearance of this scale, 
however, can be connected
to the dark matter halos. 
Indeed, the cross-correlation
of the \Lya flux decrement
and the position of halos
yields the noise cross-spectrum
of the same order of magnitude,
which is $\sim 10\%$ of the total
error 
power spectrum of the halos, see Fig.~\ref{fig:lyaxqso_pk}. 
(Here we use halos of all masses 
available 
in the simulation.)
This suggests that 
the \Lya ``shot noise''
is inherited from the halos. 
Specifically, the \Lya stochasticity 
appears consistent with the picture where 
cold gas is accumulated around massive halos,
but pushed out from the light halos 
on scales larger than their virial radii~\cite{Bolton:2016bfs}. 
The line-of-sight 
stochasticity appears to be driven 
by the non-perturbative inflow velocities.

Importantly, 
while the \Lya noise power spectrum 
has the theoretically 
expected power-law 
momentum scaling 
on large scales,
its slope becomes
shallower around $k\approx 0.6~\hMpc$, where 
it starts making a sizable 
contribution to the
total \Lya power. 
This behavior suggests 
a breakdown of the gradient 
expansion for the stochastic
component. 
At face value, 
modeling 
this non-trivial scale-dependence of the stochastic noise may 
pose a serious challenge
to the EFT-based cosmological inference.
However, in a companion paper, we show
that the flattening of the 
noise power spectrum mirrors 
the behavior of dark matter
power spectrum in redshift space, 
which suggest that this
noise might be captured 
if the EFT expansion is 
applied to the full non-linear 
matter field from an N-body simulation.\footnote{In the context of galaxy clustering such an approach
is known as the hybrid EFT~(HEFT;~\cite{Modi:2019qbt,Hadzhiyska:2021xbv, 2021MNRAS.505.1422K,Sullivan_Chen_LPNG_HEFT}).}
We leave this prospective 
line of research for future 
exploration. 

\textit{Summary and discussion.}---In this \textit{Letter}, we present the first
analytic forward model for the \Lya forest flux decrement that works successfully 
at the field level. 
Our approach bridges the gap between theory and simulation by enabling efficient modeling of the flux field directly, rather than only its summary statistics. 
We find excellent agreement between our model and the Sherwood hydrodynamic simulations for the \Lya forest and halo fields on quasi-linear scales: we reproduce the \Lya power spectrum at the 5\%-level up to $k \simlt 0.6 \hMpcinv$ and achieve accurate counts-in-cells down to $\sim 1-2 \hinvMpc$
cell radii. We achieve somewhat better results for the \Lya--halo cross-spectrum, reaching scales up to $k \simlt 1 \hMpcinv$ with a similar result for the counts-in-cells.

Our work opens up 
opportunities for multiple
high-impact contributions 
to \Lya cosmology. 
Our model can be used to bridge
the \Lya forest maps 
on large and small scales by 
analytically generating 
the large-scale models whose 
properties match the 
small-volume high-resolution 
simulations (e.g.~Illustris TNG~\cite{2017MNRAS.465.3291W, Pillepich:2017jle}, ACCEL$^2$ \cite{Chabanier:2024knr}, CAMELS \cite{CAMELS_presentation, CAMELS_DR1}, PRIYA \cite{Bird:2023evb}).
This will remove the computational bottleneck of large-volume hydrodynamic simulations and 
enable efficient simulation-based inference from the \Lya fields. 
In analogy with 
galaxy clustering, this can be done in multiple 
ways: using the 
field-level EFT inference~\cite{Nguyen:2020hxe,Nguyen:2024yth,Akitsu_FLI},
hybrid EFT-simulation-based-inference approaches~\cite{Obuljen:2022cjo,Modi:2023drt,Ivanov:2024hgq},
or building 
emulators for the \Lya forest
power spectra and bispectra~\cite{Hahn:2023kky}.
Another application
is the generation of semi-analytic 
simulations
to estimate covariance matrices,
which are urgently needed for joint analyses of the \Lya forest 
and quasars from DESI. 
Our approach can also be used for precision measurements of the 
\Lya EFT parameters at the field level without sample variance,
thus enabling EFT-based analyses 
with 
simulation-based priors 
along the lines of~\cite{Akitsu:2023eqa,Ivanov:2024hgq,Ivanov:2024xgb,Akitsu:2024lyt,Ivanov:2024dgv,Sullivan:2025eei,Akitsu_IA_FL}.
Ultimately, our approach significantly 
advances the theoretical understanding of the \Lya forest and 
its connection to the underlying matter distribution, bringing us closer to unraveling the puzzles of structure formation in the high-redshift Universe.

\textit{Acknowledgments.}---We thank Andrei Cuceu, Vid Ir\v{s}i\v{c}, Andreu Font-Ribera, Pat McDonald, and  Mike Toomey
for useful discussions, 
and Jamie Bolton, 
Jon{\'a}s Chaves-Montero, Jahmour Givans, and Andreu Font-Ribera for providing
the Sherwood files. 
This research used resources of the National Energy Research Scientific Computing Center (NERSC), a U.S. Department of Energy Office of Science User Facility operated under Contract No. DE-AC02-05CH11231.
This work was performed in part at Aspen Center for Physics, which is supported by National Science Foundation grant PHY-2210452.
JMS acknowledges that support for this work was provided by The Brinson Foundation through a Brinson Prize.
KA acknowledges supports from Fostering Joint International Research (B) under Contract No. 21KK0050 and the Japan Society for the Promotion of Science (JSPS) KAKENHI Grant No. JP24K17056. SC acknowledges support from the National Science Foundation at the IAS through NSF-BSF 2207583.


\bibliographystyle{JHEP}
\bibliography{short.bib, references}

\providecommand{\href}[2]{#2}\begingroup\raggedright\begin{thebibliography}{100}

\bibitem{McQuinn:2016}
M.~{McQuinn}, \emph{{The Evolution of the Intergalactic Medium}}, \href{https://doi.org/10.1146/annurev-astro-082214-122355}{\emph{\araa} {\bfseries 54} (2016) 313} [\href{https://arxiv.org/abs/1512.00086}{{\ttfamily 1512.00086}}].

\bibitem{Seljak:2005}
U.~{Seljak}, A.~{Makarov}, P.~{McDonald}, S.~F. {Anderson}, N.~A. {Bahcall}, J.~{Brinkmann} et~al., \emph{{Cosmological parameter analysis including SDSS Ly{\ensuremath{\alpha}} forest and galaxy bias: Constraints on the primordial spectrum of fluctuations, neutrino mass, and dark energy}}, \href{https://doi.org/10.1103/PhysRevD.71.103515}{\emph{\prd} {\bfseries 71} (2005) 103515} [\href{https://arxiv.org/abs/astro-ph/0407372}{{\ttfamily astro-ph/0407372}}].

\bibitem{Viel:2005}
M.~{Viel}, J.~{Lesgourgues}, M.~G. {Haehnelt}, S.~{Matarrese} and A.~{Riotto}, \emph{{Constraining warm dark matter candidates including sterile neutrinos and light gravitinos with WMAP and the Lyman-{\ensuremath{\alpha}} forest}}, \href{https://doi.org/10.1103/PhysRevD.71.063534}{\emph{\prd} {\bfseries 71} (2005) 063534} [\href{https://arxiv.org/abs/astro-ph/0501562}{{\ttfamily astro-ph/0501562}}].

\bibitem{McDonald06}
P.~{McDonald}, U.~{Seljak}, S.~{Burles}, D.~J. {Schlegel}, D.~H. {Weinberg}, R.~{Cen} et~al., \emph{{The Ly{\ensuremath{\alpha}} Forest Power Spectrum from the Sloan Digital Sky Survey}}, \href{https://doi.org/10.1086/444361}{\emph{\apjs} {\bfseries 163} (2006) 80} [\href{https://arxiv.org/abs/astro-ph/0405013}{{\ttfamily astro-ph/0405013}}].

\bibitem{PYB13}
N.~{Palanque-Delabrouille}, C.~{Y{\`e}che}, A.~{Borde}, J.-M. {Le Goff}, G.~{Rossi}, M.~{Viel} et~al., \emph{{The one-dimensional Ly{\ensuremath{\alpha}} forest power spectrum from BOSS}}, \href{https://doi.org/10.1051/0004-6361/201322130}{\emph{\aap} {\bfseries 559} (2013) A85} [\href{https://arxiv.org/abs/1306.5896}{{\ttfamily 1306.5896}}].

\bibitem{Chabanier:2019}
S.~{Chabanier}, N.~{Palanque-Delabrouille}, C.~{Y{\`e}che}, J.-M. {Le Goff}, E.~{Armengaud}, J.~{Bautista} et~al., \emph{{The one-dimensional power spectrum from the SDSS DR14 Ly{\ensuremath{\alpha}} forests}}, \href{https://doi.org/10.1088/1475-7516/2019/07/017}{\emph{\jcap} {\bfseries 2019} (2019) 017} [\href{https://arxiv.org/abs/1812.03554}{{\ttfamily 1812.03554}}].

\bibitem{Pedersen:2020}
C.~Pedersen, A.~Font-Ribera, K.~K. Rogers, P.~McDonald, H.~V. Peiris, A.~Pontzen et~al., \emph{{An emulator for the Lyman-$\alpha$ forest in beyond-$\Lambda$CDM cosmologies}}, \href{https://doi.org/10.1088/1475-7516/2021/05/033}{\emph{JCAP} {\bfseries 05} (2021) 033} [\href{https://arxiv.org/abs/2011.15127}{{\ttfamily 2011.15127}}].

\bibitem{2023MNRAS.526.5118R}
C.~{Ravoux}, M.~L. {Abdul Karim}, E.~{Armengaud}, M.~{Walther}, N.~G. {Kara{\c{c}}ayl{\i}}, P.~{Martini} et~al., \emph{{The Dark Energy Spectroscopic Instrument: one-dimensional power spectrum from first Ly {\ensuremath{\alpha}} forest samples with Fast Fourier Transform}}, \href{https://doi.org/10.1093/mnras/stad3008}{\emph{\mnras} {\bfseries 526} (2023) 5118} [\href{https://arxiv.org/abs/2306.06311}{{\ttfamily 2306.06311}}].

\bibitem{2024MNRAS.tmp..176K}
N.~G. {Kara{\c{c}}ayl{\i}}, P.~{Martini}, J.~{Guy}, C.~{Ravoux}, M.~L.~A. {Karim}, E.~{Armengaud} et~al., \emph{{Optimal 1D Ly{\ensuremath{\alpha}} Forest Power Spectrum Estimation - III. DESI early data}}, \href{https://doi.org/10.1093/mnras/stae171}{\emph{\mnras} (2024) } [\href{https://arxiv.org/abs/2306.06316}{{\ttfamily 2306.06316}}].

\bibitem{Viel:2010}
M.~{Viel}, M.~G. {Haehnelt} and V.~{Springel}, \emph{{The effect of neutrinos on the matter distribution as probed by the intergalactic medium}}, \href{https://doi.org/10.1088/1475-7516/2010/06/015}{\emph{\jcap} {\bfseries 2010} (2010) 015} [\href{https://arxiv.org/abs/1003.2422}{{\ttfamily 1003.2422}}].

\bibitem{Palanque2020}
N.~{Palanque-Delabrouille}, C.~{Y{\`e}che}, N.~{Sch{\"o}neberg}, J.~{Lesgourgues}, M.~{Walther}, S.~{Chabanier} et~al., \emph{{Hints, neutrino bounds, and WDM constraints from SDSS DR14 Lyman-{\ensuremath{\alpha}} and Planck full-survey data}}, \href{https://doi.org/10.1088/1475-7516/2020/04/038}{\emph{\jcap} {\bfseries 2020} (2020) 038} [\href{https://arxiv.org/abs/1911.09073}{{\ttfamily 1911.09073}}].

\bibitem{Ivanov:2024jtl}
M.~M. Ivanov, M.~W. Toomey and N.~G. Kara\c{c}ayl\i{}, \emph{{Fundamental Physics with the Lyman-Alpha Forest: Constraints on the Growth of Structure and Neutrino Masses from SDSS with Effective Field Theory}}, \href{https://doi.org/10.1103/PhysRevLett.134.091001}{\emph{Phys. Rev. Lett.} {\bfseries 134} (2025) 091001} [\href{https://arxiv.org/abs/2405.13208}{{\ttfamily 2405.13208}}].

\bibitem{He:2023oke}
A.~He, R.~An, M.~M. Ivanov and V.~Gluscevic, \emph{{Self-Interacting Neutrinos in Light of Large-Scale Structure Data}},  \href{https://arxiv.org/abs/2309.03956}{{\ttfamily 2309.03956}}.

\bibitem{He:2025jwp}
A.~He, M.~M. Ivanov, S.~Bird, R.~An and V.~Gluscevic, \emph{{A Fresh Look at Neutrino Self-Interactions With the Lyman-$\alpha$ Forest: Constraints from EFT and PRIYA}},  \href{https://arxiv.org/abs/2503.15592}{{\ttfamily 2503.15592}}.

\bibitem{Afshordi:2003}
N.~{Afshordi}, P.~{McDonald} and D.~N. {Spergel}, \emph{{Primordial Black Holes as Dark Matter: The Power Spectrum and Evaporation of Early Structures}}, \href{https://doi.org/10.1086/378763}{\emph{\apjl} {\bfseries 594} (2003) L71} [\href{https://arxiv.org/abs/astro-ph/0302035}{{\ttfamily astro-ph/0302035}}].

\bibitem{Murgia:2019}
R.~Murgia, G.~Scelfo, M.~Viel and A.~Raccanelli, \emph{{Lyman-\ensuremath{\alpha} Forest Constraints on Primordial Black Holes as Dark Matter}}, \href{https://doi.org/10.1103/PhysRevLett.123.071102}{\emph{Phys. Rev. Lett.} {\bfseries 123} (2019) 071102} [\href{https://arxiv.org/abs/1903.10509}{{\ttfamily 1903.10509}}].

\bibitem{Viel:2013}
M.~{Viel}, G.~D. {Becker}, J.~S. {Bolton} and M.~G. {Haehnelt}, \emph{{Warm dark matter as a solution to the small scale crisis: New constraints from high redshift Lyman-{\ensuremath{\alpha}} forest data}}, \href{https://doi.org/10.1103/PhysRevD.88.043502}{\emph{\prd} {\bfseries 88} (2013) 043502} [\href{https://arxiv.org/abs/1306.2314}{{\ttfamily 1306.2314}}].

\bibitem{Baur:2016}
J.~{Baur}, N.~{Palanque-Delabrouille}, C.~{Y{\`e}che}, C.~{Magneville} and M.~{Viel}, \emph{{Lyman-alpha forests cool warm dark matter}}, \href{https://doi.org/10.1088/1475-7516/2016/08/012}{\emph{\jcap} {\bfseries 2016} (2016) 012} [\href{https://arxiv.org/abs/1512.01981}{{\ttfamily 1512.01981}}].

\bibitem{Irsic17}
V.~{Ir{\v s}i{\v c}}, M.~{Viel}, M.~G. {Haehnelt}, J.~S. {Bolton}, S.~{Cristiani}, G.~D. {Becker} et~al., \emph{{New Constraints on the free-streaming of warm dark matter from intermediate and small scale Lyman-$\alpha$ forest data}}, {\emph{ArXiv e-prints} (2017) } [\href{https://arxiv.org/abs/1702.01764}{{\ttfamily 1702.01764}}].

\bibitem{Kobayashi:2017}
T.~{Kobayashi}, R.~{Murgia}, A.~{De Simone}, V.~{Ir{\v{s}}i{\v{c}}} and M.~{Viel}, \emph{{Lyman-{\ensuremath{\alpha}} constraints on ultralight scalar dark matter: Implications for the early and late universe}}, \href{https://doi.org/10.1103/PhysRevD.96.123514}{\emph{\prd} {\bfseries 96} (2017) 123514} [\href{https://arxiv.org/abs/1708.00015}{{\ttfamily 1708.00015}}].

\bibitem{Armengaud:2017}
E.~{Armengaud}, N.~{Palanque-Delabrouille}, C.~{Y{\`e}che}, D.~J.~E. {Marsh} and J.~{Baur}, \emph{{Constraining the mass of light bosonic dark matter using SDSS Lyman-{\ensuremath{\alpha}} forest}}, \href{https://doi.org/10.1093/mnras/stx1870}{\emph{\mnras} {\bfseries 471} (2017) 4606} [\href{https://arxiv.org/abs/1703.09126}{{\ttfamily 1703.09126}}].

\bibitem{Murgia:2018}
R.~{Murgia}, V.~{Ir{\v{s}}i{\v{c}}} and M.~{Viel}, \emph{{Novel constraints on noncold, nonthermal dark matter from Lyman-{\ensuremath{\alpha}} forest data}}, \href{https://doi.org/10.1103/PhysRevD.98.083540}{\emph{\prd} {\bfseries 98} (2018) 083540} [\href{https://arxiv.org/abs/1806.08371}{{\ttfamily 1806.08371}}].

\bibitem{Garzilli:2019}
A.~{Garzilli}, A.~{Magalich}, T.~{Theuns}, C.~S. {Frenk}, C.~{Weniger}, O.~{Ruchayskiy} et~al., \emph{{The Lyman-{\ensuremath{\alpha}} forest as a diagnostic of the nature of the dark matter}}, \href{https://doi.org/10.1093/mnras/stz2188}{\emph{\mnras} {\bfseries 489} (2019) 3456} [\href{https://arxiv.org/abs/1809.06585}{{\ttfamily 1809.06585}}].

\bibitem{Irsic:2020}
V.~{Ir{\v{s}}i{\v{c}}}, H.~{Xiao} and M.~{McQuinn}, \emph{{Early structure formation constraints on the ultralight axion in the postinflation scenario}}, \href{https://doi.org/10.1103/PhysRevD.101.123518}{\emph{\prd} {\bfseries 101} (2020) 123518} [\href{https://arxiv.org/abs/1911.11150}{{\ttfamily 1911.11150}}].

\bibitem{Rogers:2022}
K.~K. {Rogers}, C.~{Dvorkin} and H.~V. {Peiris}, \emph{{Limits on the Light Dark Matter-Proton Cross Section from Cosmic Large-Scale Structure}}, \href{https://doi.org/10.1103/PhysRevLett.128.171301}{\emph{\prl} {\bfseries 128} (2022) 171301} [\href{https://arxiv.org/abs/2111.10386}{{\ttfamily 2111.10386}}].

\bibitem{Villasenor:2023}
B.~{Villasenor}, B.~{Robertson}, P.~{Madau} and E.~{Schneider}, \emph{{New constraints on warm dark matter from the Lyman-{\ensuremath{\alpha}} forest power spectrum}}, \href{https://doi.org/10.1103/PhysRevD.108.023502}{\emph{\prd} {\bfseries 108} (2023) 023502} [\href{https://arxiv.org/abs/2209.14220}{{\ttfamily 2209.14220}}].

\bibitem{Irsic:2023}
V.~{Ir{\v{s}}i{\v{c}}}, M.~{Viel}, M.~G. {Haehnelt}, J.~S. {Bolton}, M.~{Molaro}, E.~{Puchwein} et~al., \emph{{Unveiling Dark Matter free-streaming at the smallest scales with high redshift Lyman-alpha forest}}, \href{https://doi.org/10.48550/arXiv.2309.04533}{\emph{arXiv e-prints} (2023) arXiv:2309.04533} [\href{https://arxiv.org/abs/2309.04533}{{\ttfamily 2309.04533}}].

\bibitem{Zaldarriaga:2002}
M.~{Zaldarriaga}, \emph{{Searching for Fluctuations in the Intergalactic Medium Temperature Using the Ly{\ensuremath{\alpha}} Forest}}, \href{https://doi.org/10.1086/324212}{\emph{\apj} {\bfseries 564} (2002) 153} [\href{https://arxiv.org/abs/astro-ph/0102205}{{\ttfamily astro-ph/0102205}}].

\bibitem{Meiksin:2009}
A.~A. {Meiksin}, \emph{{The physics of the intergalactic medium}}, \href{https://doi.org/10.1103/RevModPhys.81.1405}{\emph{Reviews of Modern Physics} {\bfseries 81} (2009) 1405} [\href{https://arxiv.org/abs/0711.3358}{{\ttfamily 0711.3358}}].

\bibitem{Viel:2006}
M.~{Viel}, J.~{Lesgourgues}, M.~G. {Haehnelt}, S.~{Matarrese} and A.~{Riotto}, \emph{{Can Sterile Neutrinos Be Ruled Out as Warm Dark Matter Candidates?}}, \href{https://doi.org/10.1103/PhysRevLett.97.071301}{\emph{\prl} {\bfseries 97} (2006) 071301} [\href{https://arxiv.org/abs/astro-ph/0605706}{{\ttfamily astro-ph/0605706}}].

\bibitem{Walther:2019}
M.~{Walther}, J.~{O{\~n}orbe}, J.~F. {Hennawi} and Z.~{Luki{\'c}}, \emph{{New Constraints on IGM Thermal Evolution from the Ly{\ensuremath{\alpha}} Forest Power Spectrum}}, \href{https://doi.org/10.3847/1538-4357/aafad1}{\emph{\apj} {\bfseries 872} (2019) 13} [\href{https://arxiv.org/abs/1808.04367}{{\ttfamily 1808.04367}}].

\bibitem{Bolton:2008}
J.~S. {Bolton}, M.~{Viel}, T.~S. {Kim}, M.~G. {Haehnelt} and R.~F. {Carswell}, \emph{{Possible evidence for an inverted temperature-density relation in the intergalactic medium from the flux distribution of the Ly{\ensuremath{\alpha}} forest}}, \href{https://doi.org/10.1111/j.1365-2966.2008.13114.x}{\emph{\mnras} {\bfseries 386} (2008) 1131} [\href{https://arxiv.org/abs/0711.2064}{{\ttfamily 0711.2064}}].

\bibitem{Garzilli:2012}
A.~{Garzilli}, J.~S. {Bolton}, T.~S. {Kim}, S.~{Leach} and M.~{Viel}, \emph{{The intergalactic medium thermal history at redshift z = 1.7-3.2 from the Ly{\ensuremath{\alpha}} forest: a comparison of measurements using wavelets and the flux distribution}}, \href{https://doi.org/10.1111/j.1365-2966.2012.21223.x}{\emph{\mnras} {\bfseries 424} (2012) 1723} [\href{https://arxiv.org/abs/1202.3577}{{\ttfamily 1202.3577}}].

\bibitem{Gaikwad:2019}
P.~{Gaikwad}, R.~{Srianand}, V.~{Khaire} and T.~R. {Choudhury}, \emph{{Effect of non-equilibrium ionization on derived physical conditions of the high-z intergalactic medium}}, \href{https://doi.org/10.1093/mnras/stz2692}{\emph{\mnras} {\bfseries 490} (2019) 1588} [\href{https://arxiv.org/abs/1812.01016}{{\ttfamily 1812.01016}}].

\bibitem{Boera:2019}
E.~{Boera}, G.~D. {Becker}, J.~S. {Bolton} and F.~{Nasir}, \emph{{Revealing Reionization with the Thermal History of the Intergalactic Medium: New Constraints from the Ly{\ensuremath{\alpha}} Flux Power Spectrum}}, \href{https://doi.org/10.3847/1538-4357/aafee4}{\emph{\apj} {\bfseries 872} (2019) 101} [\href{https://arxiv.org/abs/1809.06980}{{\ttfamily 1809.06980}}].

\bibitem{Gaikwad:2021}
P.~{Gaikwad}, R.~{Srianand}, M.~G. {Haehnelt} and T.~R. {Choudhury}, \emph{{A consistent and robust measurement of the thermal state of the IGM at 2 {\ensuremath{\leq}} z {\ensuremath{\leq}} 4 from a large sample of Ly {\ensuremath{\alpha}} forest spectra: evidence for late and rapid He II reionization}}, \href{https://doi.org/10.1093/mnras/stab2017}{\emph{\mnras} {\bfseries 506} (2021) 4389} [\href{https://arxiv.org/abs/2009.00016}{{\ttfamily 2009.00016}}].

\bibitem{Wilson:2022}
B.~{Wilson}, V.~{Ir{\v{s}}i{\v{c}}} and M.~{McQuinn}, \emph{{A measurement of the Ly {\ensuremath{\beta}} forest power spectrum and its cross with the Ly {\ensuremath{\alpha}} forest in X-Shooter XQ-100}}, \href{https://doi.org/10.1093/mnras/stab3017}{\emph{\mnras} {\bfseries 509} (2022) 2423} [\href{https://arxiv.org/abs/2106.04837}{{\ttfamily 2106.04837}}].

\bibitem{Villasenor:2022}
B.~{Villasenor}, B.~{Robertson}, P.~{Madau} and E.~{Schneider}, \emph{{Inferring the Thermal History of the Intergalactic Medium from the Properties of the Hydrogen and Helium Ly{\ensuremath{\alpha}} Forest}}, \href{https://doi.org/10.3847/1538-4357/ac704e}{\emph{\apj} {\bfseries 933} (2022) 59} [\href{https://arxiv.org/abs/2111.00019}{{\ttfamily 2111.00019}}].

\bibitem{Goldstein:2023gnw}
S.~Goldstein, J.~C. Hill, V.~Ir\v{s}i\v{c} and B.~D. Sherwin, \emph{{Canonical Hubble-Tension-Resolving Early Dark Energy Cosmologies are Inconsistent with the Lyman-$\alpha$ Forest}},  \href{https://arxiv.org/abs/2303.00746}{{\ttfamily 2303.00746}}.

\bibitem{Garny:2018byk}
M.~Garny, T.~Konstandin, L.~Sagunski and S.~Tulin, \emph{{Lyman-$\alpha$ forest constraints on interacting dark sectors}}, \href{https://doi.org/10.1088/1475-7516/2018/09/011}{\emph{JCAP} {\bfseries 09} (2018) 011} [\href{https://arxiv.org/abs/1805.12203}{{\ttfamily 1805.12203}}].

\bibitem{Fuss:2022zyt}
L.~Fu\ss{} and M.~Garny, \emph{{Decaying Dark Matter and Lyman-$\alpha$ forest constraints}},  \href{https://arxiv.org/abs/2210.06117}{{\ttfamily 2210.06117}}.

\bibitem{Seljak:2006bg}
U.~Seljak, A.~Slosar and P.~McDonald, \emph{{Cosmological parameters from combining the Lyman-alpha forest with CMB, galaxy clustering and SN constraints}}, \href{https://doi.org/10.1088/1475-7516/2006/10/014}{\emph{JCAP} {\bfseries 10} (2006) 014} [\href{https://arxiv.org/abs/astro-ph/0604335}{{\ttfamily astro-ph/0604335}}].

\bibitem{McDonald:2007}
P.~{McDonald} and D.~J. {Eisenstein}, \emph{{Dark energy and curvature from a future baryonic acoustic oscillation survey using the Lyman-{\ensuremath{\alpha}} forest}}, \href{https://doi.org/10.1103/PhysRevD.76.063009}{\emph{\prd} {\bfseries 76} (2007) 063009} [\href{https://arxiv.org/abs/astro-ph/0607122}{{\ttfamily astro-ph/0607122}}].

\bibitem{Slosar2013}
A.~{Slosar}, V.~{Ir{\v{s}}i{\v{c}}}, D.~{Kirkby}, S.~{Bailey}, N.~G. {Busca}, T.~{Delubac} et~al., \emph{{Measurement of baryon acoustic oscillations in the Lyman-{\ensuremath{\alpha}} forest fluctuations in BOSS data release 9}}, \href{https://doi.org/10.1088/1475-7516/2013/04/026}{\emph{\jcap} {\bfseries 2013} (2013) 026} [\href{https://arxiv.org/abs/1301.3459}{{\ttfamily 1301.3459}}].

\bibitem{Busca:2013}
N.~G. {Busca}, T.~{Delubac}, J.~{Rich}, S.~{Bailey}, A.~{Font-Ribera}, D.~{Kirkby} et~al., \emph{{Baryon acoustic oscillations in the Ly{\ensuremath{\alpha}} forest of BOSS quasars}}, \href{https://doi.org/10.1051/0004-6361/201220724}{\emph{\aap} {\bfseries 552} (2013) A96} [\href{https://arxiv.org/abs/1211.2616}{{\ttfamily 1211.2616}}].

\bibitem{dMdB:2020}
H.~{du Mas des Bourboux}, J.~{Rich}, A.~{Font-Ribera}, V.~{de Sainte Agathe}, J.~{Farr}, T.~{Etourneau} et~al., \emph{{The Completed SDSS-IV Extended Baryon Oscillation Spectroscopic Survey: Baryon Acoustic Oscillations with Ly{\ensuremath{\alpha}} Forests}}, \href{https://doi.org/10.3847/1538-4357/abb085}{\emph{\apj} {\bfseries 901} (2020) 153} [\href{https://arxiv.org/abs/2007.08995}{{\ttfamily 2007.08995}}].

\bibitem{Cuceu:2021}
A.~{Cuceu}, A.~{Font-Ribera}, B.~{Joachimi} and S.~{Nadathur}, \emph{{Cosmology beyond BAO from the 3D distribution of the Lyman-{\ensuremath{\alpha}} forest}}, \href{https://doi.org/10.1093/mnras/stab1999}{\emph{\mnras} {\bfseries 506} (2021) 5439} [\href{https://arxiv.org/abs/2103.14075}{{\ttfamily 2103.14075}}].

\bibitem{Cuceu:2023}
A.~{Cuceu}, A.~{Font-Ribera}, S.~{Nadathur}, B.~{Joachimi} and P.~{Martini}, \emph{{Constraints on the Cosmic Expansion Rate at Redshift 2.3 from the Lyman-{\ensuremath{\alpha}} Forest}}, \href{https://doi.org/10.1103/PhysRevLett.130.191003}{\emph{\prl} {\bfseries 130} (2023) 191003} [\href{https://arxiv.org/abs/2209.13942}{{\ttfamily 2209.13942}}].

\bibitem{Gordon:2023}
C.~{Gordon}, A.~{Cuceu}, J.~{Chaves-Montero}, A.~{Font-Ribera}, A.~{Xochitl Gonz{\'a}lez-Morales}, J.~{Aguilar} et~al., \emph{{3D Correlations in the Lyman-$\alpha$ Forest from Early DESI Data}}, \href{https://doi.org/10.48550/arXiv.2308.10950}{\emph{arXiv e-prints} (2023) arXiv:2308.10950} [\href{https://arxiv.org/abs/2308.10950}{{\ttfamily 2308.10950}}].

\bibitem{DESI:2025zgx}
{\scshape DESI} collaboration, \emph{{DESI DR2 Results II: Measurements of Baryon Acoustic Oscillations and Cosmological Constraints}},  \href{https://arxiv.org/abs/2503.14738}{{\ttfamily 2503.14738}}.

\bibitem{DESI:2016}
{DESI Collaboration}, A.~{Aghamousa}, J.~{Aguilar}, S.~{Ahlen}, S.~{Alam}, L.~E. {Allen} et~al., \emph{{The DESI Experiment Part I: Science,Targeting, and Survey Design}}, {\emph{arXiv e-prints} (2016) arXiv:1611.00036} [\href{https://arxiv.org/abs/1611.00036}{{\ttfamily 1611.00036}}].

\bibitem{DESI:2022}
B.~{Abareshi}, J.~{Aguilar}, S.~{Ahlen}, S.~{Alam}, D.~M. {Alexander}, R.~{Alfarsy} et~al., \emph{{Overview of the Instrumentation for the Dark Energy Spectroscopic Instrument}}, {\emph{arXiv e-prints} (2022) arXiv:2205.10939} [\href{https://arxiv.org/abs/2205.10939}{{\ttfamily 2205.10939}}].

\bibitem{DESI_BAO_2024}
{DESI Collaboration}, A.~G. {Adame}, J.~{Aguilar}, S.~{Ahlen}, S.~{Alam}, D.~M. {Alexander} et~al., \emph{{DESI 2024 VI: Cosmological Constraints from the Measurements of Baryon Acoustic Oscillations}}, \href{https://doi.org/10.48550/arXiv.2404.03002}{\emph{arXiv e-prints} (2024) arXiv:2404.03002} [\href{https://arxiv.org/abs/2404.03002}{{\ttfamily 2404.03002}}].

\bibitem{DESI_lya_2024}
{DESI Collaboration}, A.~G. {Adame}, J.~{Aguilar}, S.~{Ahlen}, S.~{Alam}, D.~M. {Alexander} et~al., \emph{{DESI 2024 IV: Baryon Acoustic Oscillations from the Lyman Alpha Forest}}, \href{https://doi.org/10.48550/arXiv.2404.03001}{\emph{arXiv e-prints} (2024) arXiv:2404.03001} [\href{https://arxiv.org/abs/2404.03001}{{\ttfamily 2404.03001}}].

\bibitem{Schlegel:22_spec_roadmap}
D.~J. {Schlegel}, S.~{Ferraro}, G.~{Aldering}, C.~{Baltay}, S.~{BenZvi}, R.~{Besuner} et~al., \emph{{A Spectroscopic Road Map for Cosmic Frontier: DESI, DESI-II, Stage-5}}, \href{https://doi.org/10.48550/arXiv.2209.03585}{\emph{arXiv e-prints} (2022) arXiv:2209.03585} [\href{https://arxiv.org/abs/2209.03585}{{\ttfamily 2209.03585}}].

\bibitem{Besuner25:spec_s5}
R.~{Besuner}, A.~{Dey}, A.~{Drlica-Wagner}, H.~{Ebina}, G.~{Fernandez Moroni}, S.~{Ferraro} et~al., \emph{{The Spectroscopic Stage-5 Experiment}}, \href{https://doi.org/10.48550/arXiv.2503.07923}{\emph{arXiv e-prints} (2025) arXiv:2503.07923} [\href{https://arxiv.org/abs/2503.07923}{{\ttfamily 2503.07923}}].

\bibitem{Font-Ribera:2018}
A.~{Font-Ribera}, P.~{McDonald} and A.~{Slosar}, \emph{{How to estimate the 3D power spectrum of the Lyman-{\ensuremath{\alpha}} forest}}, \href{https://doi.org/10.1088/1475-7516/2018/01/003}{\emph{\jcap} {\bfseries 2018} (2018) 003} [\href{https://arxiv.org/abs/1710.11036}{{\ttfamily 1710.11036}}].

\bibitem{deBelsunce:2024knf}
R.~de~Belsunce, O.~H.~E. Philcox, V.~Irsic, P.~McDonald, J.~Guy and N.~Palanque-Delabrouille, \emph{{The 3D Lyman-\ensuremath{\alpha} forest power spectrum from eBOSS DR16}}, \href{https://doi.org/10.1093/mnras/stae2035}{\emph{Mon. Not. Roy. Astron. Soc.} {\bfseries 533} (2024) 3756} [\href{https://arxiv.org/abs/2403.08241}{{\ttfamily 2403.08241}}].

\bibitem{Nguyen:2024yth}
N.-M. Nguyen, F.~Schmidt, B.~Tucci, M.~Reinecke and A.~Kosti{\'c}, \emph{{How Much Information Can Be Extracted from Galaxy Clustering at the Field Level?}}, \href{https://doi.org/10.1103/PhysRevLett.133.221006}{\emph{Phys. Rev. Lett.} {\bfseries 133} (2024) 221006} [\href{https://arxiv.org/abs/2403.03220}{{\ttfamily 2403.03220}}].

\bibitem{Akitsu_FLI}
K.~Akitsu, M.~Simonovi\'c, S.-F. Chen, G.~Cabass and M.~Zaldarriaga, \emph{Cosmology inference with perturbative forward modeling at the field level: comparison with the joint power spectrum and bispectrum analysis}, .

\bibitem{Hadzhiyska:2023}
B.~{Hadzhiyska}, A.~{Font-Ribera}, A.~{Cuceu}, S.~{Chabanier}, J.~{Aguilar}, D.~{Brooks} et~al., \emph{{Planting a Lyman alpha forest on ABACUSSUMMIT}}, \href{https://doi.org/10.1093/mnras/stad1920}{\emph{\mnras} {\bfseries 524} (2023) 1008} [\href{https://arxiv.org/abs/2305.08899}{{\ttfamily 2305.08899}}].

\bibitem{Hadzhiyska:2025cvk}
B.~Hadzhiyska, R.~de~Belsunce, A.~Cuceu, J.~Guy, M.~M. Ivanov, H.~Coquinot et~al., \emph{{Measuring and unbiasing the BAO shift in the Lyman-Alpha forest with AbacusSummit}},  \href{https://arxiv.org/abs/2503.13442}{{\ttfamily 2503.13442}}.

\bibitem{Croft98}
R.~A.~C. {Croft}, \emph{{Characterization of Lyman Alpha Spectra and Predictions of Structure Formation Models: A Flux Statistics Approach}},  in \emph{Eighteenth Texas Symposium on Relativistic Astrophysics}, A.~V. {Olinto}, J.~A. {Frieman} and D.~N. {Schramm}, eds., p.~664, 1998, \href{https://arxiv.org/abs/astro-ph/9701166}{{\ttfamily astro-ph/9701166}}.

\bibitem{2022ApJ...930..109Q}
M.~{Qezlou}, A.~B. {Newman}, G.~C. {Rudie} and S.~{Bird}, \emph{{Characterizing Protoclusters and Protogroups at z 2.5 Using Ly{\ensuremath{\alpha}} Tomography}}, \href{https://doi.org/10.3847/1538-4357/ac6259}{\emph{\apj} {\bfseries 930} (2022) 109} [\href{https://arxiv.org/abs/2112.03930}{{\ttfamily 2112.03930}}].

\bibitem{Peirani:2014}
S.~{Peirani}, D.~H. {Weinberg}, S.~{Colombi}, J.~{Blaizot}, Y.~{Dubois} and C.~{Pichon}, \emph{{LyMAS: Predicting Large-scale Ly{\ensuremath{\alpha}} Forest Statistics from the Dark Matter Density Field}}, \href{https://doi.org/10.1088/0004-637X/784/1/11}{\emph{\apj} {\bfseries 784} (2014) 11} [\href{https://arxiv.org/abs/1306.1533}{{\ttfamily 1306.1533}}].

\bibitem{Peirani:2022}
S.~{Peirani}, S.~{Prunet}, S.~{Colombi}, C.~{Pichon}, D.~H. {Weinberg}, C.~{Laigle} et~al., \emph{{LyMAS reloaded: improving the predictions of the large-scale Lyman-{\ensuremath{\alpha}} forest statistics from dark matter density and velocity fields}}, \href{https://doi.org/10.1093/mnras/stac1344}{\emph{\mnras} {\bfseries 514} (2022) 3222} [\href{https://arxiv.org/abs/2204.06365}{{\ttfamily 2204.06365}}].

\bibitem{Sorini:2016}
D.~{Sorini}, J.~{O{\~n}orbe}, Z.~{Luki{\'c}} and J.~F. {Hennawi}, \emph{{Modeling the Ly{\ensuremath{\alpha}} Forest in Collisionless Simulations}}, \href{https://doi.org/10.3847/0004-637X/827/2/97}{\emph{\apj} {\bfseries 827} (2016) 97} [\href{https://arxiv.org/abs/1602.08099}{{\ttfamily 1602.08099}}].

\bibitem{2022ApJ...927..230S}
F.~{Sinigaglia}, F.-S. {Kitaura}, A.~{Balaguera-Antol{\'\i}nez}, I.~{Shimizu}, K.~{Nagamine}, M.~{S{\'a}nchez-Benavente} et~al., \emph{{Mapping the Three-dimensional Ly{\ensuremath{\alpha}} Forest Large-scale Structure in Real and Redshift Space}}, \href{https://doi.org/10.3847/1538-4357/ac5112}{\emph{\apj} {\bfseries 927} (2022) 230}.

\bibitem{2024A&A...682A..21S}
F.~{Sinigaglia}, F.~S. {Kitaura}, K.~{Nagamine}, Y.~{Oku} and A.~{Balaguera-Antol{\'\i}nez}, \emph{{Field-level Lyman-{\ensuremath{\alpha}} forest modeling in redshift space via augmented nonlocal Fluctuating Gunn-Peterson Approximation}}, \href{https://doi.org/10.1051/0004-6361/202346931}{\emph{\aap} {\bfseries 682} (2024) A21} [\href{https://arxiv.org/abs/2305.10428}{{\ttfamily 2305.10428}}].

\bibitem{Jacobus:2024yev}
C.~Jacobus, S.~Chabanier, P.~Harrington, J.~Emberson, Z.~Luki\'c and S.~Habib, \emph{{A Gigaparsec-Scale Hydrodynamic Volume Reconstructed with Deep Learning}},  \href{https://arxiv.org/abs/2411.16920}{{\ttfamily 2411.16920}}.

\bibitem{McDonald:2009dh}
P.~McDonald and A.~Roy, \emph{{Clustering of dark matter tracers: generalizing bias for the coming era of precision LSS}}, \href{https://doi.org/10.1088/1475-7516/2009/08/020}{\emph{JCAP} {\bfseries 0908} (2009) 020} [\href{https://arxiv.org/abs/0902.0991}{{\ttfamily 0902.0991}}].

\bibitem{Baumann:2010tm}
D.~Baumann, A.~Nicolis, L.~Senatore and M.~Zaldarriaga, \emph{{Cosmological Non-Linearities as an Effective Fluid}}, \href{https://doi.org/10.1088/1475-7516/2012/07/051}{\emph{JCAP} {\bfseries 1207} (2012) 051} [\href{https://arxiv.org/abs/1004.2488}{{\ttfamily 1004.2488}}].

\bibitem{Carrasco:2013mua}
J.~J.~M. Carrasco, S.~Foreman, D.~Green and L.~Senatore, \emph{{The Effective Field Theory of Large Scale Structures at Two Loops}}, \href{https://doi.org/10.1088/1475-7516/2014/07/057}{\emph{JCAP} {\bfseries 07} (2014) 057} [\href{https://arxiv.org/abs/1310.0464}{{\ttfamily 1310.0464}}].

\bibitem{Ivanov:2022mrd}
M.~M. Ivanov, \emph{{Effective Field Theory for Large Scale Structure}},  \href{https://arxiv.org/abs/2212.08488}{{\ttfamily 2212.08488}}.

\bibitem{Schmittfull:2020trd}
M.~Schmittfull, M.~Simonovi\'c, M.~M. Ivanov, O.~H.~E. Philcox and M.~Zaldarriaga, \emph{{Modeling Galaxies in Redshift Space at the Field Level}}, \href{https://doi.org/10.1088/1475-7516/2021/05/059}{\emph{JCAP} {\bfseries 05} (2021) 059} [\href{https://arxiv.org/abs/2012.03334}{{\ttfamily 2012.03334}}].

\bibitem{Obuljen:2022cjo}
A.~Obuljen, M.~Simonovi\'c, A.~Schneider and R.~Feldmann, \emph{{Modeling HI at the field level}}, \href{https://doi.org/10.1103/PhysRevD.108.083528}{\emph{Phys. Rev. D} {\bfseries 108} (2023) 083528} [\href{https://arxiv.org/abs/2207.12398}{{\ttfamily 2207.12398}}].

\bibitem{McDonald:1999dt}
P.~McDonald, J.~Miralda-Escude, M.~Rauch, W.~L.~W. Sargent, T.~A. Barlow, R.~Cen et~al., \emph{{The Observed probability distribution function, power spectrum, and correlation function of the transmitted flux in the Lyman-alpha forest}}, \href{https://doi.org/10.1086/317079}{\emph{Astrophys. J.} {\bfseries 543} (2000) 1} [\href{https://arxiv.org/abs/astro-ph/9911196}{{\ttfamily astro-ph/9911196}}].

\bibitem{Givans:2020sez}
J.~J. Givans and C.~M. Hirata, \emph{{Redshift-space streaming velocity effects on the Lyman-$\alpha$ forest baryon acoustic oscillation scale}}, \href{https://doi.org/10.1103/PhysRevD.102.023515}{\emph{Phys. Rev. D} {\bfseries 102} (2020) 023515} [\href{https://arxiv.org/abs/2002.12296}{{\ttfamily 2002.12296}}].

\bibitem{Desjacques:2018pfv}
V.~Desjacques, D.~Jeong and F.~Schmidt, \emph{{The Galaxy Power Spectrum and Bispectrum in Redshift Space}}, \href{https://doi.org/10.1088/1475-7516/2018/12/035}{\emph{JCAP} {\bfseries 1812} (2018) 035} [\href{https://arxiv.org/abs/1806.04015}{{\ttfamily 1806.04015}}].

\bibitem{Chen:2021rnb}
S.-F. Chen, Z.~Vlah and M.~White, \emph{{The Ly$\alpha$ forest flux correlation function: a perturbation theory perspective}}, \href{https://doi.org/10.1088/1475-7516/2021/05/053}{\emph{JCAP} {\bfseries 05} (2021) 053} [\href{https://arxiv.org/abs/2103.13498}{{\ttfamily 2103.13498}}].

\bibitem{Ivanov:2023yla}
M.~M. Ivanov, \emph{{Lyman alpha forest power spectrum in effective field theory}}, \href{https://doi.org/10.1103/PhysRevD.109.023507}{\emph{Phys. Rev. D} {\bfseries 109} (2024) 023507} [\href{https://arxiv.org/abs/2309.10133}{{\ttfamily 2309.10133}}].

\bibitem{Belsunce_Sullivan_skewspectrum}
R.~de~Belsunce, J.~M. Sullivan and P.~McDonald, ``The lyman-$\alpha$ forest skew spectrum.'' 2025.

\bibitem{Bolton:2016bfs}
J.~S. Bolton, E.~Puchwein, D.~Sijacki, M.~G. Haehnelt, T.-S. Kim, A.~Meiksin et~al., \emph{{The Sherwood simulation suite: overview and data comparisons with the Lyman \ensuremath{\alpha} forest at redshifts 2 \ensuremath{\leq} z \ensuremath{\leq} 5}}, \href{https://doi.org/10.1093/mnras/stw2397}{\emph{Mon. Not. Roy. Astron. Soc.} {\bfseries 464} (2017) 897} [\href{https://arxiv.org/abs/1605.03462}{{\ttfamily 1605.03462}}].

\bibitem{McDonald:2001}
P.~McDonald, \emph{{Toward a measurement of the cosmological geometry at Z 2: predicting lyman-alpha forest correlation in three dimensions, and the potential of future data sets}}, \href{https://doi.org/10.1086/345945}{\emph{Astrophys. J.} {\bfseries 585} (2003) 34} [\href{https://arxiv.org/abs/astro-ph/0108064}{{\ttfamily astro-ph/0108064}}].

\bibitem{Kaiser:1987qv}
N.~Kaiser, \emph{{Clustering in real space and in redshift space}}, {\emph{Mon. Not. Roy. Astron. Soc.} {\bfseries 227} (1987) 1}.

\bibitem{Schmittfull:2018yuk}
M.~Schmittfull, M.~Simonović, V.~Assassi and M.~Zaldarriaga, \emph{{Modeling Biased Tracers at the Field Level}}, \href{https://doi.org/10.1103/PhysRevD.100.043514}{\emph{Phys.\ Rev.\ D} {\bfseries 100} (2019) 043514} [\href{https://arxiv.org/abs/1811.10640}{{\ttfamily 1811.10640}}].

\bibitem{Chen:2021}
S.-F. Chen, Z.~Vlah and M.~White, \emph{{The Ly$\alpha$ forest flux correlation function: a perturbation theory perspective}}, \href{https://doi.org/10.1088/1475-7516/2021/05/053}{\emph{JCAP} {\bfseries 05} (2021) 053} [\href{https://arxiv.org/abs/2103.13498}{{\ttfamily 2103.13498}}].

\bibitem{Senatore:2014via}
L.~Senatore and M.~Zaldarriaga, \emph{{The IR-resummed Effective Field Theory of Large Scale Structures}}, \href{https://doi.org/10.1088/1475-7516/2015/02/013}{\emph{JCAP} {\bfseries 1502} (2015) 013} [\href{https://arxiv.org/abs/1404.5954}{{\ttfamily 1404.5954}}].

\bibitem{Baldauf:2015xfa}
T.~Baldauf, M.~Mirbabayi, M.~Simonović and M.~Zaldarriaga, \emph{{Equivalence Principle and the Baryon Acoustic Peak}}, \href{https://doi.org/10.1103/PhysRevD.92.043514}{\emph{Phys. Rev.} {\bfseries D92} (2015) 043514} [\href{https://arxiv.org/abs/1504.04366}{{\ttfamily 1504.04366}}].

\bibitem{Vlah:2015zda}
Z.~Vlah, U.~Seljak, M.~Y. Chu and Y.~Feng, \emph{{Perturbation theory, effective field theory, and oscillations in the power spectrum}}, \href{https://doi.org/10.1088/1475-7516/2016/03/057}{\emph{JCAP} {\bfseries 1603} (2016) 057} [\href{https://arxiv.org/abs/1509.02120}{{\ttfamily 1509.02120}}].

\bibitem{Blas:2015qsi}
D.~Blas, M.~Garny, M.~M. Ivanov and S.~Sibiryakov, \emph{{Time-Sliced Perturbation Theory for Large Scale Structure I: General Formalism}}, \href{https://doi.org/10.1088/1475-7516/2016/07/052}{\emph{JCAP} {\bfseries 1607} (2016) 052} [\href{https://arxiv.org/abs/1512.05807}{{\ttfamily 1512.05807}}].

\bibitem{Blas:2016sfa}
D.~Blas, M.~Garny, M.~M. Ivanov and S.~Sibiryakov, \emph{{Time-Sliced Perturbation Theory II: Baryon Acoustic Oscillations and Infrared Resummation}}, \href{https://doi.org/10.1088/1475-7516/2016/07/028}{\emph{JCAP} {\bfseries 1607} (2016) 028} [\href{https://arxiv.org/abs/1605.02149}{{\ttfamily 1605.02149}}].

\bibitem{Ivanov:2018gjr}
M.~M. Ivanov and S.~Sibiryakov, \emph{{Infrared Resummation for Biased Tracers in Redshift Space}}, \href{https://doi.org/10.1088/1475-7516/2018/07/053}{\emph{JCAP} {\bfseries 1807} (2018) 053} [\href{https://arxiv.org/abs/1804.05080}{{\ttfamily 1804.05080}}].

\bibitem{Vasudevan:2019ewf}
A.~Vasudevan, M.~M. Ivanov, S.~Sibiryakov and J.~Lesgourgues, \emph{{Time-sliced perturbation theory with primordial non-Gaussianity and effects of large bulk flows on inflationary oscillating features}}, \href{https://doi.org/10.1088/1475-7516/2019/09/037}{\emph{JCAP} {\bfseries 09} (2019) 037} [\href{https://arxiv.org/abs/1906.08697}{{\ttfamily 1906.08697}}].

\bibitem{Ivanov:2024xgb}
M.~M. Ivanov, A.~Obuljen, C.~Cuesta-Lazaro and M.~W. Toomey, \emph{{Full-shape analysis with simulation-based priors: cosmological parameters and the structure growth anomaly}},  \href{https://arxiv.org/abs/2409.10609}{{\ttfamily 2409.10609}}.

\bibitem{Givans2022}
J.~J. {Givans}, A.~{Font-Ribera}, A.~{Slosar}, L.~{Seeyave}, C.~{Pedersen}, K.~K. {Rogers} et~al., \emph{{Non-linearities in the Lyman-{\ensuremath{\alpha}} forest and in its cross-correlation with dark matter halos}}, \href{https://doi.org/10.1088/1475-7516/2022/09/070}{\emph{\jcap} {\bfseries 2022} (2022) 070} [\href{https://arxiv.org/abs/2205.00962}{{\ttfamily 2205.00962}}].

\bibitem{Springel05}
V.~{Springel}, \emph{{The cosmological simulation code GADGET-2}}, \href{https://doi.org/10.1111/j.1365-2966.2005.09655.x}{\emph{\mnras} {\bfseries 364} (2005) 1105} [\href{https://arxiv.org/abs/astro-ph/0505010}{{\ttfamily astro-ph/0505010}}].

\bibitem{Planck_first}
{Planck Collaboration}, P.~A.~R. {Ade}, N.~{Aghanim}, C.~{Armitage-Caplan}, M.~{Arnaud}, M.~{Ashdown} et~al., \emph{{Planck 2013 results. XVI. Cosmological parameters}}, \href{https://doi.org/10.1051/0004-6361/201321591}{\emph{\aap} {\bfseries 571} (2014) A16} [\href{https://arxiv.org/abs/1303.5076}{{\ttfamily 1303.5076}}].

\bibitem{Chudaykin:2025gsh}
A.~{Chudaykin} and M.~M. {Ivanov}, \emph{{Lyman Alpha Forest - Halo Cross-Correlations in Effective Field Theory}}, {\emph{arXiv e-prints} (2025) arXiv:2501.04770} [\href{https://arxiv.org/abs/2501.04770}{{\ttfamily 2501.04770}}].

\bibitem{Cieplak:2015kra}
A.~M. Cieplak and A.~Slosar, \emph{{Towards physics responsible for large-scale Lyman-\ensuremath{\alpha} forest bias parameters}}, \href{https://doi.org/10.1088/1475-7516/2016/03/016}{\emph{JCAP} {\bfseries 03} (2016) 016} [\href{https://arxiv.org/abs/1509.07875}{{\ttfamily 1509.07875}}].

\bibitem{Givans:2022qgb}
J.~J. Givans, A.~Font-Ribera, A.~Slosar, L.~Seeyave, C.~Pedersen, K.~K. Rogers et~al., \emph{{Non-linearities in the Lyman-\ensuremath{\alpha} forest and in its cross-correlation with dark matter halos}}, \href{https://doi.org/10.1088/1475-7516/2022/09/070}{\emph{JCAP} {\bfseries 09} (2022) 070} [\href{https://arxiv.org/abs/2205.00962}{{\ttfamily 2205.00962}}].

\bibitem{Modi:2019qbt}
C.~Modi, S.-F. Chen and M.~White, \emph{{Simulations and symmetries}}, \href{https://doi.org/10.1093/mnras/staa251}{\emph{Mon. Not. Roy. Astron. Soc.} {\bfseries 492} (2020) 5754} [\href{https://arxiv.org/abs/1910.07097}{{\ttfamily 1910.07097}}].

\bibitem{Hadzhiyska:2021xbv}
B.~Hadzhiyska, C.~Garc{\'\i}a-Garc{\'\i}a, D.~Alonso, A.~Nicola and A.~Slosar, \emph{{Hefty enhancement of cosmological constraints from the DES Y1 data using a hybrid effective field theory approach to galaxy bias}}, \href{https://doi.org/10.1088/1475-7516/2021/09/020}{\emph{JCAP} {\bfseries 09} (2021) 020} [\href{https://arxiv.org/abs/2103.09820}{{\ttfamily 2103.09820}}].

\bibitem{2021MNRAS.505.1422K}
N.~{Kokron}, J.~{DeRose}, S.-F. {Chen}, M.~{White} and R.~H. {Wechsler}, \emph{{The cosmology dependence of galaxy clustering and lensing from a hybrid N-body-perturbation theory model}}, \href{https://doi.org/10.1093/mnras/stab1358}{\emph{\mnras} {\bfseries 505} (2021) 1422} [\href{https://arxiv.org/abs/2101.11014}{{\ttfamily 2101.11014}}].

\bibitem{Sullivan_Chen_LPNG_HEFT}
J.~M. {Sullivan} and S.-F. {Chen}, \emph{{Local primordial non-Gaussian bias at the field level}}, \href{https://doi.org/10.1088/1475-7516/2025/03/016}{\emph{\jcap} {\bfseries 2025} (2025) 016} [\href{https://arxiv.org/abs/2410.18039}{{\ttfamily 2410.18039}}].

\bibitem{2017MNRAS.465.3291W}
R.~{Weinberger}, V.~{Springel}, L.~{Hernquist}, A.~{Pillepich}, F.~{Marinacci}, R.~{Pakmor} et~al., \emph{{Simulating galaxy formation with black hole driven thermal and kinetic feedback}}, \href{https://doi.org/10.1093/mnras/stw2944}{\emph{\mnras} {\bfseries 465} (2017) 3291} [\href{https://arxiv.org/abs/1607.03486}{{\ttfamily 1607.03486}}].

\bibitem{Pillepich:2017jle}
A.~Pillepich et~al., \emph{{Simulating Galaxy Formation with the IllustrisTNG Model}}, \href{https://doi.org/10.1093/mnras/stx2656}{\emph{Mon. Not. Roy. Astron. Soc.} {\bfseries 473} (2018) 4077} [\href{https://arxiv.org/abs/1703.02970}{{\ttfamily 1703.02970}}].

\bibitem{Chabanier:2024knr}
S.~Chabanier, C.~Ravoux, L.~Latrille, J.~Sexton, E.~Armengaud, J.~Bautista et~al., \emph{{The ACCEL2 project: simulating Lyman-\ensuremath{\alpha} forest in large-volume hydrodynamical simulations}}, \href{https://doi.org/10.1093/mnras/stae2255}{\emph{Mon. Not. Roy. Astron. Soc.} {\bfseries 534} (2024) 2674} [\href{https://arxiv.org/abs/2407.04473}{{\ttfamily 2407.04473}}].

\bibitem{CAMELS_presentation}
F.~{Villaescusa-Navarro}, D.~{Angl{\'e}s-Alc{\'a}zar}, S.~{Genel}, D.~N. {Spergel}, R.~S. {Somerville}, R.~{Dave} et~al., \emph{{The CAMELS Project: Cosmology and Astrophysics with Machine-learning Simulations}}, \href{https://doi.org/10.3847/1538-4357/abf7ba}{\emph{\apj} {\bfseries 915} (2021) 71} [\href{https://arxiv.org/abs/2010.00619}{{\ttfamily 2010.00619}}].

\bibitem{CAMELS_DR1}
F.~{Villaescusa-Navarro}, S.~{Genel}, D.~{Angl{\'e}s-Alc{\'a}zar}, L.~A. {Perez}, P.~{Villanueva-Domingo}, D.~{Wadekar} et~al., \emph{{The CAMELS Project: Public Data Release}}, \href{https://doi.org/10.3847/1538-4365/acbf47}{\emph{\apjs} {\bfseries 265} (2023) 54} [\href{https://arxiv.org/abs/2201.01300}{{\ttfamily 2201.01300}}].

\bibitem{Bird:2023evb}
S.~Bird, M.~Fernandez, M.-F. Ho, M.~Qezlou, R.~Monadi, Y.~Ni et~al., \emph{{PRIYA: a new suite of Lyman-{\ensuremath{\alpha}} forest simulations for cosmology}}, \href{https://doi.org/10.1088/1475-7516/2023/10/037}{\emph{JCAP} {\bfseries 10} (2023) 037} [\href{https://arxiv.org/abs/2306.05471}{{\ttfamily 2306.05471}}].

\bibitem{Nguyen:2020hxe}
N.-M. Nguyen, F.~Schmidt, G.~Lavaux and J.~Jasche, \emph{{Impacts of the physical data model on the forward inference of initial conditions from biased tracers}}, \href{https://doi.org/10.1088/1475-7516/2021/03/058}{\emph{JCAP} {\bfseries 03} (2021) 058} [\href{https://arxiv.org/abs/2011.06587}{{\ttfamily 2011.06587}}].

\bibitem{Modi:2023drt}
C.~Modi and O.~H.~E. Philcox, \emph{{Hybrid SBI or How I Learned to Stop Worrying and Learn the Likelihood}},  \href{https://arxiv.org/abs/2309.10270}{{\ttfamily 2309.10270}}.

\bibitem{Ivanov:2024hgq}
M.~M. Ivanov, C.~Cuesta-Lazaro, S.~Mishra-Sharma, A.~Obuljen and M.~W. Toomey, \emph{{Full-shape analysis with simulation-based priors: constraints on single field inflation from BOSS}},  \href{https://arxiv.org/abs/2402.13310}{{\ttfamily 2402.13310}}.

\bibitem{Hahn:2023kky}
C.~Hahn, M.~Eickenberg, S.~Ho, J.~Hou, P.~Lemos, E.~Massara et~al., \emph{{${\rm SIMBIG}$: The First Cosmological Constraints from the Non-Linear Galaxy Bispectrum}},  \href{https://arxiv.org/abs/2310.15243}{{\ttfamily 2310.15243}}.

\bibitem{Akitsu:2023eqa}
K.~Akitsu, Y.~Li and T.~Okumura, \emph{{Quadratic shape biases in three-dimensional halo intrinsic alignments}}, \href{https://doi.org/10.1088/1475-7516/2023/08/068}{\emph{JCAP} {\bfseries 08} (2023) 068} [\href{https://arxiv.org/abs/2306.00969}{{\ttfamily 2306.00969}}].

\bibitem{Akitsu:2024lyt}
K.~Akitsu, \emph{{Mapping the galaxy-halo connection to the galaxy bias: implication to the HOD-informed prior}},  \href{https://arxiv.org/abs/2410.08998}{{\ttfamily 2410.08998}}.

\bibitem{Ivanov:2024dgv}
M.~M. Ivanov et~al., \emph{{The Millennium and Astrid galaxies in effective field theory: comparison with galaxy-halo connection models at the field level}},  \href{https://arxiv.org/abs/2412.01888}{{\ttfamily 2412.01888}}.

\bibitem{Sullivan:2025eei}
J.~M. Sullivan, C.~Cuesta-Lazaro, M.~M. Ivanov, Y.~Ni, S.~Bose, B.~Hadzhiyska et~al., \emph{{High-redshift Millennium and Astrid galaxies in effective field theory at the field level}},  \href{https://arxiv.org/abs/2505.03626}{{\ttfamily 2505.03626}}.

\bibitem{Akitsu_IA_FL}
K.~Akitsu, Y.~Li and T.~Okumura, ``Modeling the intrinsic alignment at the field level.'' 2025.

\bibitem{Bernardeau:2001CPT}
F.~Bernardeau, S.~Colombi, E.~Gaztanaga and R.~Scoccimarro, \emph{{Large scale structure of the universe and cosmological perturbation theory}}, \href{https://doi.org/10.1016/S0370-1573(02)00135-7}{\emph{\physrep} {\bfseries 367} (2002) 1} [\href{https://arxiv.org/abs/astro-ph/0112551}{{\ttfamily astro-ph/0112551}}].

\end{thebibliography}\endgroup

\newpage 

\pagebreak
\widetext
\begin{center}
\textbf{\large Supplemental Material}
\end{center}
\setcounter{equation}{0}
\setcounter{figure}{0}
\setcounter{table}{0}
\setcounter{page}{1}
\makeatletter
\renewcommand{\theequation}{S\arabic{equation}}
\renewcommand{\thefigure}{S\arabic{figure}}
\renewcommand{\thetable}{S\arabic{table}}

\textit{Details of the forward model.}---We now introduce the EFT theory model, briefly summarizing Ref.~\cite{Ivanov:2023yla}, to which we refer the reader for a fuller presentation. The \Lya forest flux decrement, $\delta_F=F/\bar F-1$, can be described perturbatively up to cubic order using the following, generalized form
\be
\begin{split}
 \delta^{(s)}_F(\k)= &\sum_{n=1}^3 \Big[ \prod_{j=1}^n\int_{}\frac{d^3{\bf k}_j}{(2\pi)^3} \delta^{(1)}(\k_j)\Big]
 K_n(\k_1,...,\k_n)\times (2\pi)^3\delta^{(3)}_D(\k-\k_1-...-\k_j)\,,
 \end{split}
\ee
where $f$ is the logarithmic growth factor, $\mu\equiv k_\parallel/k$ with the cosine of the angle
to the line-of-sight, $\td_D$ the Dirac delta function, $\td^{(1)}$ the linear density field and $K_n$ are the perturbative kernels in redshift space (for galaxies these are denoted $Z$; see e.g.~\cite{Bernardeau:2001CPT} for a review)  
\be
\label{eq:K2full}
\begin{split}
& K_1(\k) = b_1-b_\eta f\mu^2\,,\\
& K_2(\k_1,\k_2)=\frac{b_2}{2}+b_{\mathcal{G}_2}\left(\frac{(\k_1\cdot \k_2)^2}{k_1^2 k_2^2}-1\right)+b_1F_2(\k_1,\k_2)  -b_\eta f\mu^2 G_2(\k_1,\k_2) - fb_{\delta \eta}\frac{\mu_2^2+\mu_1^2}{2} +b_{\eta^2}f^2\mu_1^2\mu_2^2\\
& +b_1f\frac{\mu_1\mu_2}{2}\left(\frac{k_2}{k_1} + \frac{k_1}{k_2}\right)
-b_\eta f^2\frac{\mu_1\mu_2}{2}\left(\frac{k_2}{k_1}\mu_2^2 + \frac{k_1}{k_2}\mu_1^2\right) + b_{(KK)_\parallel}\left(\mu_1\mu_2 \frac{(\k_1\cdot \k_2)}{k_1k_2}
-\frac{\mu_1^2+\mu_2^2}{3}+\frac{1}{9}
\right)\\
& + b_{\Pi^{(2)}_\parallel}\left(\mu_1\mu_2 \frac{(\k_1\cdot \k_2)}{k_1k_2}+\frac{5}{7}\mu^2 \left(1-\frac{(\k_1\cdot \k_2)^2}{k_1^2 k_2^2}\right)\right)\,,
\end{split} 
\ee
where $K_3$ is given in Eq.~(4.16) in Ref.~\cite{Ivanov:2023yla} and  $F_2$ and $G_2$ are the density and velocity kernels from standard cosmological perturbation theory \cite{Bernardeau:2001CPT}, respectively. Each quadratic operator above can be written as 
\be 
\mathcal{O}(\k) = \int_\p F_\mathcal{O}(\k_1,\k_2)\delta_1(\k-\p)\delta_1(\p)\,,
\ee 
where $\int_\p = \int \frac{\mathrm{d}^3p}{(2\pi)^3}$. At linear order the free coefficients (or bias parameters) are $\{b_1,\,b_\eta\}$ with six additional parameters at quadratic order $\{b_2\equiv b_{\delta^2},b_{\mathcal{G}_2},b_{(KK)_\parallel},b_{\delta\eta},b_{\eta^2},b_{\Pi^{[2]}_\parallel}\}$ and an additional five parameters at cubic order $\{b_{\Gamma_3},b_{\delta\Pi^{[2]}_\parallel},b_{\eta\Pi^{[2]}_\parallel},b_{(K\Pi^{[2]})_\parallel},b_{\Pi^{[3]}_\parallel}\}$ in the linear density field $\td^{(1)}$. 
Following the nomenclature of Eq.~\eqref{eq:general_expansion} we denote the bias parameters as $b_{\mathcal{O}}$. 

The cubic forward model
is obtained by 
promoting the linear and 
quadratic 
bias parameters to transfer functions. In this case all cubic
operators from $K_3$ are absorbed
into the linear transfer
function $\beta_1$.
Note that the contribution of the ${\Pi^{(2)}_\parallel}$  
operator can be completely absorbed 
into other operators. 
Equivalently, the contribution of ${\Pi^{(2)}_\parallel}$
vanishes as a result of the Gram-Schmidt
orthogonalization process~\cite{Schmittfull:2018yuk}.
Another operator
that appears 
in our forward model is the Zel'dovich matter 
density 
in redshift space, 
\be 
\delta_Z(\k)
=\int d^3 \q~ 
e^{-i\k\cdot(\q+\vpsi(\q)+f\hat{\bm{z}}(\vpsi(\q)\cdot \hat{\bm{z}}))}~\,,
\ee 
where $\q$ is the Lagrangian (initial)
coordinate and $\hat{\bm{z}}$
is the line-of-sight
unit vector. Using
$\delta_Z$ we 
can build the 
correct 
redshift-space distortion
contributions that define the $\eta$ field.
Specifically, $\delta_Z-\frac{3}{7}f\mu^2\mathcal{G}_2$
is equivalent to $\eta$
from Eq.~\eqref{eq:K2full}
at the quadratic order,
which we use to define
the $\tilde\eta^\perp$
operator. 
This operator starts
at the quadratic order
as its linear contribution
is fully degenerate with $\delta_1$ and hence it 
vanishes upon orthogonalization. 
Finally, we also add the $\delta^3$
operator and with the appropriate transfer function,
which is needed to ensure that the 
error power spectrum 
does not receive constant contributions 
from the deterministic 
fields up to the three-loop 
power spectrum order~\cite{Schmittfull:2018yuk}.
Note that all operators in our forward model are ``shifted''
by the Zel'dovich displacement
$\vpsi$ as
\be 
\tilde{\mathcal{O}}(\k)
=\int d^3 \q~\mathcal{O}(\q)
e^{-i\k\cdot(\q+\vpsi(\q)+f\hat{\bm{n}}(\vpsi(\q)\cdot \hat{\bm{n}}))}~\,.
\ee 
In particular, this 
reproduces
contributions from the 
``shifts''
of the linear fields 
$\delta$ and $\eta$,
whose coefficients
must be fixed by the linear bias parameters $b_1$
and $b_\eta$
by the equivalence 
principle.

\begin{figure*}
    \centering
    \includegraphics[width=1\linewidth]{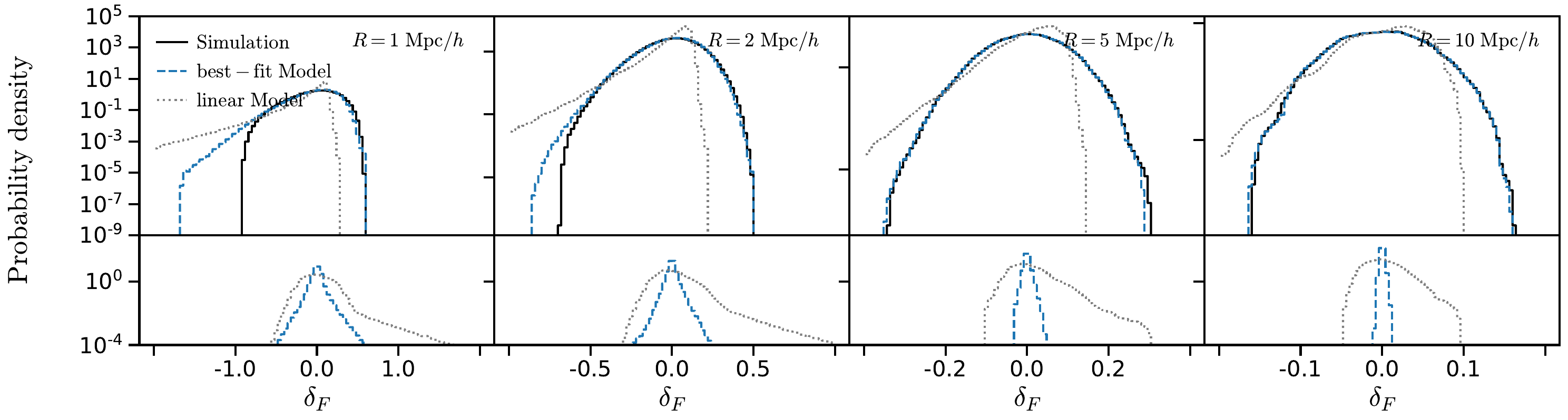}
    \caption{
    Comparison of the histogram of the flux decrement from the simulation snapshot (blue) to the best-fit cubic model (black) and the linear model (gray). The bottom panel shows the residuals. The differences between the linear forward model and the simulation are visible for \textit{all} smoothing scales. This emphasizes the importance of using non-linear bias terms and illustrates the striking breakdown of the linear model even on the largest scales. Both maps are smoothed using a 3D Gaussian isotropic kernel with radii $R=1,\, 2,\, 5,\, 10 \hinvMpc$ from left to right, respectively. }
    \label{fig:lya_pdf_1}
\end{figure*}

\begin{figure}
    \centering
    \includegraphics[width=1\linewidth]{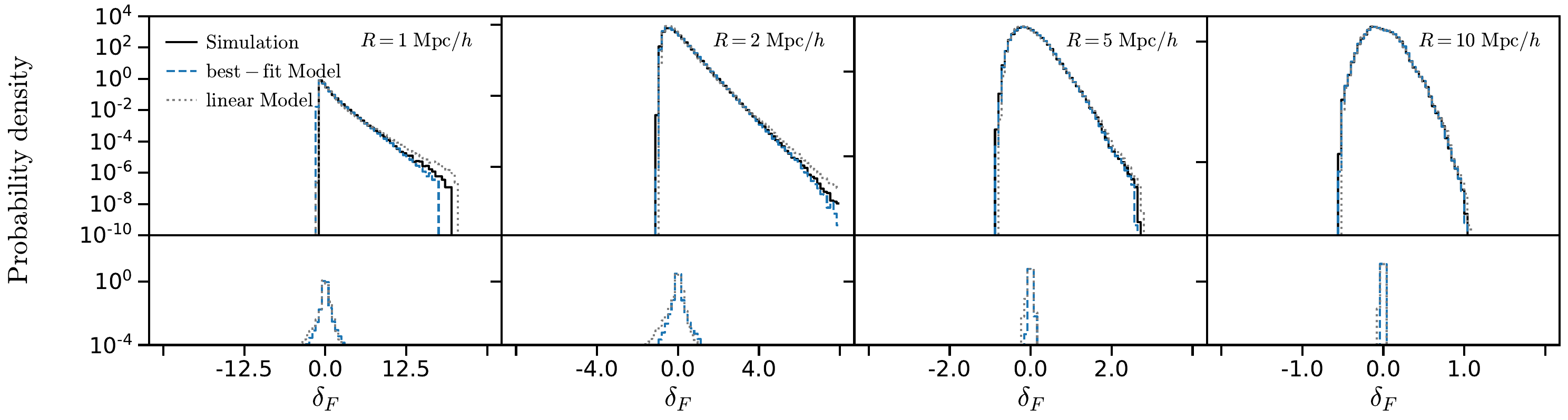}
    \caption{Same as Fig.~\ref{fig:lya_pdf_1} but for the halo field (using all available halo masses).}
    \label{fig:halo_pdf}
\end{figure}

\textit{Field-level comparison.}---In Fig.~\ref{fig:lya_pdf_1} we compare the simulated and forward modeled field by measuring the one-point probability density function (see,~\cite{Cieplak:2015kra} for a discussion in this context). Therefore, we measure a histogram of each field and apply different smoothing scales of the Gaussian isotropic kernel in the range $R=1-10 \hinvMpc$.\footnote{We emphasize that this smoothing procedure is not applicable to real data and serves for illustration purposes only, see \cite{Belsunce_Sullivan_skewspectrum} for a discussion.} Whilst for small smoothing scales $R=1,\, 2 \hinvMpc$ the tails visually appear to be large, we note that these differences are at the percent level given the number of available modes for each field. The statistical moments of the distributions in Fig.~\ref{fig:lya_pdf_1} are given in Tab.~\ref{tab:moments_pdf}, quantifying the visual agreement between the histograms. It is interesting to note that the variance is very close to zero for all smoothing scales (even in the presence of visually large tails for small smoothing scales). Following baseline expectation, the agreement between the distribution improves with increasing smoothing scale.  For a Gaussian distribution the skewness and kurtosis would vanish, thus investigating higher order, i.e.,~non-Gaussian statistics such as the bispectrum of the residuals is a fruitful avenue that we leave to future work. In Fig.~\ref{fig:halo_pdf} we show the corresponding PDF for the halo forward model (using all halo masses) and we quantify the agreement between the model and the simulation in Tab.~\ref{tab:moments_pdf_halo}. As expected the cubic forward model performs better for the \Lya forest than for halos.

\begin{table}
\centering
\begin{tabular}{cccccccccccccccc}
\hline \hline
& \multicolumn{5}{c}{Variance} 
& \multicolumn{5}{c}{Skewness} 
& \multicolumn{5}{c}{Kurtosis} \\
\cmidrule(lr){2-6} \cmidrule(lr){7-11} \cmidrule(lr){12-16}
$R$ 
& $\td_{\rm truth}$ & $\td_{\rm lin.}$ &$\td_{\rm best-fit}$ & $\Delta\td_{\rm lin.}$ & $\Delta\td$ 
& $\td_{\rm truth}$ & $\td_{\rm lin.}$ &$\td_{\rm best-fit}$ & $\Delta\td_{\rm lin.}$ & $\Delta\td$
& $\td_{\rm truth}$ & $\td_{\rm lin.}$ &$\td_{\rm best-fit}$ & $\Delta\td_{\rm lin.}$ & $\Delta\td$ \\
\hline
1  & 0.0478 & 0.0265 & 0.0456 & 0.0212 & 0.0021 & -0.4897 & -3.5673 & -0.8158 & 0.6284 & 0.2276 & 0.0271 & 22.0562 & 0.9458 & 3.4062 & 6.6934 \\
2  & 0.0234 & 0.0162 & 0.0231 & 0.0072 & 0.0004 & -0.3316 & -2.5132 & -0.4566 & 0.8553 & 0.2904 & 0.0304 & 10.5751 & 0.2591 & 2.7512 & 5.7661 \\
5  & 0.0076 & 0.0064 & 0.0076 & 0.0013 & 0.0000 & -0.0951 & -1.4013 & -0.1221 & 0.9423 & 0.2692 & -0.1209 & 3.1266 & -0.1402 & 2.0861 & 2.7730 \\
10 & 0.0027 & 0.0024 & 0.0027 & 0.0003 & 0.0000 & 0.0079 & -0.8388 & 0.0067 & 0.5289 & 0.2182 & -0.3643 & 0.9234 & -0.3736 & 0.6342 & 1.2638 \\
\hline \hline
\end{tabular}
\caption{Statistical moments (variance, skewness, and kurtosis) of the simulated ($\td_{\rm truth}$) and forward modeled ($\td_{\rm best-fit}$ for the cubic model and $\td_{\rm lin.}$ for the linear model, respectively) flux decrement fields as well as their residuals for different smoothing scales $R$ (in $\hinvMpc$). Following baseline expectation, with increasing smoothing radius $R$ the agreement between the one-point probability density functions of the forward model and the simulated field improves. Linear theory consistently under performs compared to the best-fit cubic forward model. Note that skewness and kurtosis both vanish for a Gaussian distribution.
}
\label{tab:moments_pdf}
\end{table}

\begin{table}
\centering
\begin{tabular}{cccccccccccccccc}
\hline \hline
& \multicolumn{5}{c}{Variance} 
& \multicolumn{5}{c}{Skewness} 
& \multicolumn{5}{c}{Kurtosis} \\
\cmidrule(lr){2-6} \cmidrule(lr){7-11} \cmidrule(lr){12-16}
$R$ 
& $\td_{\rm truth}$ & $\td_{\rm lin.}$ &$\td_{\rm best-fit}$ & $\Delta\td_{\rm lin.}$ & $\Delta\td$ 
& $\td_{\rm truth}$ & $\td_{\rm lin.}$ &$\td_{\rm best-fit}$ & $\Delta\td_{\rm lin.}$ & $\Delta\td$
& $\td_{\rm truth}$ & $\td_{\rm lin.}$ &$\td_{\rm best-fit}$ & $\Delta\td_{\rm lin.}$ & $\Delta\td$ \\
\hline
1  & 1.1387 & 1.0443 & 1.1036 & 0.0598 & 0.0372 & 2.5784 & 3.1960 & 2.5437 & -2.1006 & 0.3057 & 10.6183 & 17.5130 & 10.2336 & 70.4289 & 46.1026 \\
2  & 0.5201 & 0.5026 & 0.5158 & 0.0088 & 0.0043 & 1.8373 & 2.1445 & 1.8187 & -1.4736 & 0.4262 & 5.5515 & 7.6961 & 5.2637 & 25.6934 & 23.3843 \\
5  & 0.1581 & 0.1563 & 0.1578 & 0.0007 & 0.0002 & 0.9539 & 1.0875 & 0.9478 & -0.4991 & 0.2757 & 1.4494 & 1.8756 & 1.3447 & 2.5080 & 2.6281 \\
10 & 0.0542 & 0.0539 & 0.0542 & 0.0001 & 0.0000 & 0.5037 & 0.5782 & 0.5040 & -0.1707 & 0.2434 & 0.2324 & 0.3621 & 0.2074 & 0.2139 & 0.4162 \\
\hline \hline
\end{tabular}
\caption{Same as Tab.~\ref{tab:moments_pdf} but for the halo field as a function of smoothing radius, $R$, and for all halo masses with redshift space distortions applied along the $z$-axis. }
\label{tab:moments_pdf_halo}
\end{table}

\begin{figure*}
    \centering
    \includegraphics[width=1\linewidth]{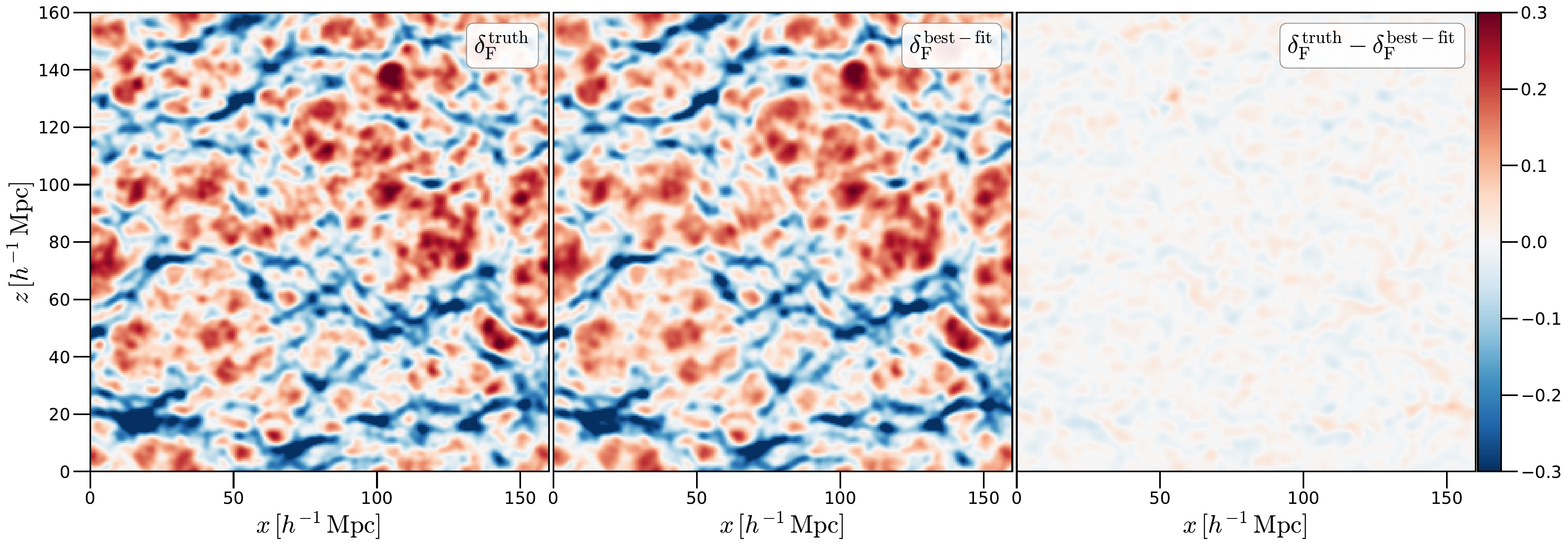}
    \includegraphics[width=1\linewidth]{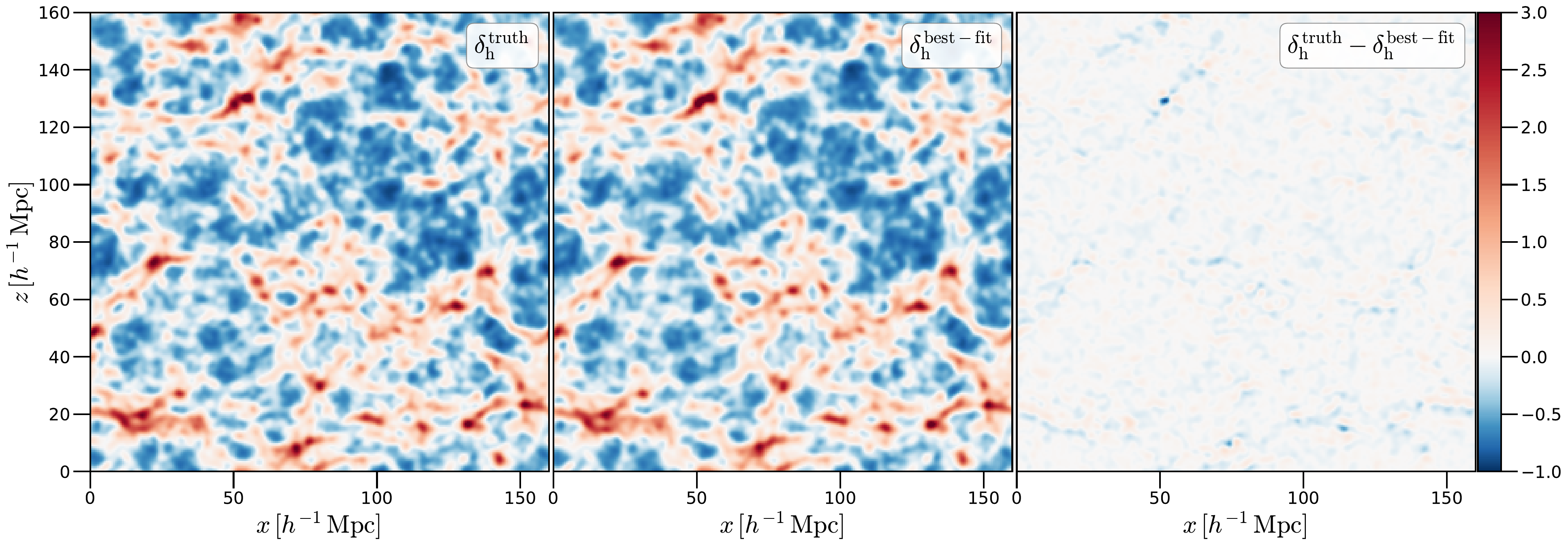}
    \caption{Same as Fig.~\ref{fig:lya_rsd_field} in the $x-z$ plane for the \Lya field (\textit{first row}) and the halo field using all available halo masses (\textit{second row}). The redshift space distortions are applied along the $z$ axis.
    }
    \label{fig:lya_rsd_field_yaxis}
\end{figure*}
In Fig.~\ref{fig:lya_rsd_field_yaxis} we show slices through the simulation snapshot, $\td^{\rm truth}$, the best-fit field from our perturbative model, $\td^{\rm best-fit}$, and the residuals in the $x-z$ plane for the \Lya (top row) and halo field (bottom row). Again, the largest residuals for the \Lya field appear in underdense regions, \textit{i.e.}~in the vicinity of halos. This $x-z$ projection visualizes the newly introduced line-of-sight operators but shows the same data as Fig.~\ref{fig:lya_rsd_field}.

\end{document}